\documentclass[notitlepage,twocolumn,prl,nobalancelastpage,amssymb,superscriptaddress,showpacs]{revtex4-1}

\usepackage{Symbols}
\usepackage{Shortcuts}
\usepackage{Text}

\usepackage{epsfig}
\usepackage{subfigure}
\usepackage[colorlinks=true,citecolor=blue,linkcolor=blue]{hyperref}
\usepackage{verbatim}

\def \ekr{{\ve_{k_R+}}}
\def \Tef{{\tilde T}}
\def \mef{{\tilde \m}}

\def \zef{{\tilde z}}
\def \bq {{\mathbf{q}}}

\renewcommand{\vec}[1]{{\bf #1}}

\begin{document}
\title{Floquet metal to insulator phase transitions in semiconductor nanowires}

\date{\today}

\author{Iliya Esin}
\affiliation{\mbox{Physics Department, Technion, 320003 Haifa, Israel}}
\author{Mark S. Rudner}
\affiliation{\mbox{Center for Quantum Devices and Niels Bohr International Academy,}
\mbox{Niels Bohr Institute, University of Copenhagen, 2100 Copenhagen, Denmark}}
\author{Netanel H. Lindner}
\affiliation{\mbox{Physics Department, Technion, 320003 Haifa, Israel}}

\begin{abstract}
 We study steady-states of semiconductor nanowires subjected to strong resonant time-periodic drives. The steady-states arise from the balance between electron-phonon scattering, electron-hole recombination via photo-emission, and Auger scattering processes. We show that tuning the strength of the driving field drives a transition between an electron-hole metal (EHM) phase and a Floquet insulator (FI) phase. We study the critical point controlling this transition. The EHM-to-FI transition can be observed by monitoring the presence of peaks in the density-density response function which are associated with the Fermi momentum of the EHM phase, and are absent in the FI phase.
Our results may help guide future studies towards inducing novel non-equilibrium phases of matter by periodic driving.
\end{abstract}

\maketitle

Coherent time-periodic driving provides a versatile tool for inducing novel properties in solid state and atomic systems~\cite{Kitagawa2010,Wang2013,Rudner2013,Rechtsman2013,Jotzu2014,Grushin2014,Mahmood2016,Titum2016,Nathan2016,Maczewsky2017,Oka2019}. 
Prominent applications include Floquet engineering of band topology, light-induced superconductivity, and ultrafast spintronics \cite{Kimel2005,Stanciu2007,Yao2007,Oka2009,Kirilyuk2010,Fausti2011,Lindner2011,Lindner2013,Katan2013,Cayssol2013,Hu2014,Mitrano2016,Cavalleri2018}.
In many contexts, interesting phenomena may be observed in the short-time dynamics of driven systems \cite{Gu2011,Kitagawa2011,DAlessio2014}.
Under appropriate conditions, at long times, Floquet systems may also exhibit nontrivial {\it steady-state} characteristics such as topological responses or time-crystalline behavior \cite{Iadecola2013,Dehghani2014,Dehghani2015,Iadecola2015,Iadecola2015a,Shirai2015,Liu2015,Dehghani2016,Else2016,Khemani2016,Yao2017,Choi2017,Zhang2017,Esin2018,McIver2018}.

In this paper we investigate transitions between distinct phases 
realized in the steady states of a periodically-driven semiconductor nanowire.
We study the case where the drive frequency is larger than the band gap of the material.
Such a system serves as a prototype for a Floquet topological insulator in which a ``resonant drive'' is used to induce an effective band inversion \cite{Lindner2011,Seetharam2015,Esin2018,Seetharam2019}.
We are interested in the regime where the steady state of the system is well-described in terms of electronic populations of the system's Floquet-Bloch states, with a nearly insulator-like filling in the Floquet basis.
Here, the Floquet bands provide a good basis for identifying the physical properties and response characteristics of the many-body steady state.

\begin{figure}[t]
  \centering
  \includegraphics[width=8.6cm]{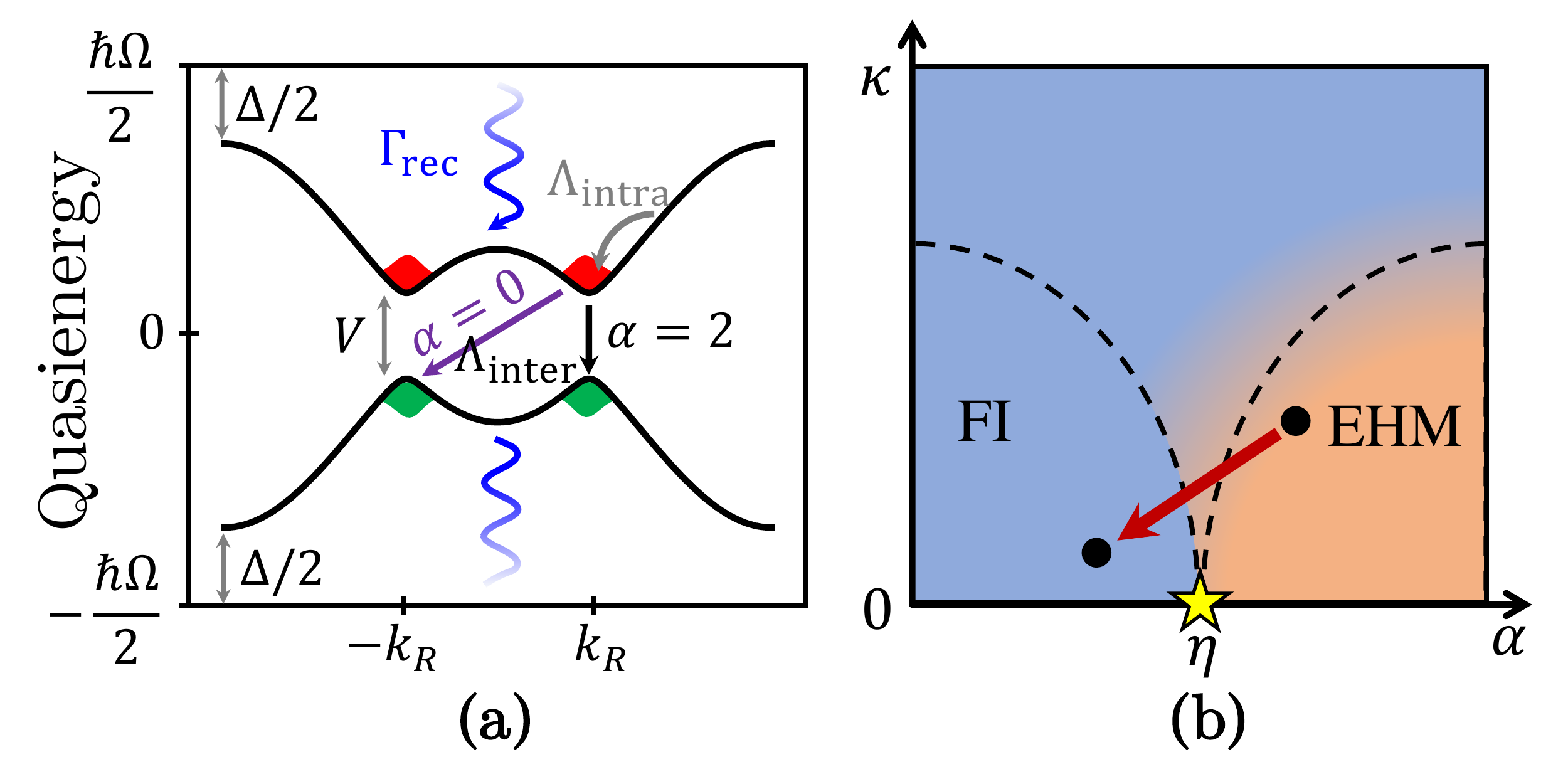}\\
  \caption{(a) Floquet spectrum of the periodically drive system described by the model in Eq.~\eqref{eq:Hamiltonian} for $M=0.4\hb\W$, $A=B=0.2\hb\W$, and $V=0.12\hb\W$. Wiggly arrow indicates photon-mediated Floquet Umklapp excitations. The black and purple straight arrows respectively describe small momentum (``vertical'') electron-hole phonon-mediated recombination, corresponding to $\a=2$, and large momentum (``diagonal'') recombination processes, corresponding to $\a=0$, respectively. The gray straight arrow indicates phonon-mediated intra-band relaxation. (b) A phase diagram of the steady-state distribution as a function of the ``balance'' parameter $\ka$, and the ``bottleneck'' parameter, $\a$. The system exhibits a quantum phase transition at $\ka\to0$ and $\a=\h$, where $\h$ is the exponent appearing in the $\w$-dependence in the density of states of phonons. For $\ka>0$, the phase transition becomes a crossover at finite effective temperature, separating an electron-hole metal (EHM) from a Floquet insulator (FI). Red arrow indicates the EHM-to-FI transition due to variation of the driving field strength.
    \label{fig:PhaseDiagram}}
\end{figure}

An ideal Floquet insulator is characterized by having a set of Floquet bands that are fully filled, while the remaining Floquet bands are empty. In a resonantly-driven system, the effective band inversion implies that such a Floquet insulator state features electronic populations in both valence and conduction band states of the non-driven system. From the point of view of the system's equilibrium band structure, the steady state therefore hosts a non-equilibrium density of excited electrons and holes. The natural relaxation of these excited electrons and holes through radiative recombination is manifested in the Floquet picture as interband transitions 
that create {\it excitations} away from the ideal Floquet insulator state (making holes in the nominally filled Floquet band, and putting electrons into the nominally empty Floquet band).
Spontaneous emission therefore contributes a source of ``quantum heating'' in the Floquet basis (see Fig.~\ref{fig:PhaseDiagram}a, wiggly arrow) \cite{Dykman2011}. Similarly, 
inter-Floquet-band transitions arising from electron-electron interactions may also create excitations away from the ideal Floquet insulator state. At the same time, spontaneous electron-phonon scattering may lead to interband transitions that {\it reduce} the number of excitations, helping to ``cool'' the system towards the Floquet insulator state (Fig.~\ref{fig:PhaseDiagram}a, straight arrows). The steady state is determined by the competition between these various scattering processes.

In this paper, we show that the electronic steady-states of resonantly-driven semiconductor nanowires may exhibit two phases: (i) an electron-hole metal (EHM) phase, which features sharp Fermi surfaces for electron and hole excitations in the nominally empty and filled Floquet bands, respectively; and (ii) a Floquet insulator (FI) phase, in which the electron and hole excitations are distributed as a non-degenerate Fermi gas in the Floquet basis. We show that the system's phase diagram is controlled by a quantum critical point, with a critical region separating the two phases, see Fig.~\ref{fig:PhaseDiagram}b. The transition between the EHM and the FI across the critical region is reminiscent of a finite-temperature crossover. The properties of the crossover are determined by the effective temperature of the electron and hole excitations in the Floquet bands. Starting from the EHM, a transition to the FI can be induced by increasing the driving field's strength beyond a critical value. We further show that the EHM phase can be experimentally identified by observing peaks in the density-density response associated with the Fermi momentum of the excited electrons. This response gives rise to Friedel oscillations in the density induced by local inhomogeneities or an external potential.



%

\section{Model for periodically-driven semiconductor nanowires}
To study the phase diagram, we use a simple tight-binding model describing a periodically driven nanowire with the Hamiltonian $\hat\cH_0(t)=\sum_{k}\hat{\tb c}_k\dg H_k(t)\hat{\tb c}_k$. Here $\hat {\tb c}_k=\mat{\hat c_{\cA,k}&\hat c_{\cB,k}}^T$ is a vector of operators annihilating fermions in two orbitals, $\ket{\cA}$ and $\ket{\cB}$, with crystal momentum $k$ along the wire. Throughout this work we neglect the spin of the electron. We write the single particle Bloch Hamiltonian $H_k(t)$ in the form:
\Eq{
H_k(t)=[M-B \cos(k a)]\s^z+ A \sin(k a)\s^y+V \cos(\W t)\s^x,
\label{eq:Hamiltonian}
}
where $\s^i$ are Pauli matrices defining an orbital basis, and $A$, $B$ and $M$ are constants. The periodic drive induces a local coupling between the orbitals, with strength $V$.
Throughout this work we consider a half-filled system, which is a band insulator in the absence of the drive. 

The Floquet eigenstates of the time-dependent problem satisfy $\bS{i\hb\frac{\dpa }{\dpa t}-H_k(t)}\ket{\y_{k\n}(t)}=0$, with $\ket{\y_{k\n}(t)}=e^{-i\ve_{k\n} t/\hb}\ket{\phi_{k\n}(t)}$. Here  $\ket{\f_{k\n}(t)} = \sum_m e^{-im\W t}\ket{\f^m_{k\n}}$ is time-periodic with period $\cT=2\pi/\Omega$, and $\ve_{k\n}$ is the quasienergy.  Throughout, we use the convention $-\hb\Omega/2\leq\ve_{k\n}<\hb\Omega/2$.

We study the regime where $2\hb\W$ is less than the total bandwidth ($2|M|+2|B|$).
In this regime, the drive only resonantly couples states in the two bands via {\it single photon} resonances; these resonances occur at crystal momentum values $k=\pm k_R$ where $\hbar\Omega$ matches the splitting between valence and conduction bands of the nondriven system.
The resulting Floquet spectrum exhibits a gap proportional to the driving field strength, $V$, separating the upper ($\n=+$) and lower ($\n=-$) Floquet bands. A plot of a generic quasienergy band-structure for the Hamiltonian in Eq.~\eqref{eq:Hamiltonian} is shown in Fig.~\ref{fig:PhaseDiagram}a.

%

In addition to the coherent effects of the drive, captured in Eq.~(\ref{eq:Hamiltonian}), we also describe the key dissipative processes that govern the steady states of the system.
To this end, we incorporate in the model couplings between the electrons of the nanowire and acoustic phonons of the $d$-dimensional substrate upon which it sits (with $d \ge 2$), as well as coupling of the electronic system to its (three-dimensional) electromagnetic environment.
The electromagnetic coupling allows for radiative recombination of electron-hole pairs via spontaneous photon emission, which provides the primary source of heating in the Floquet basis.
The effect of electron-electron interactions on the steady state is discussed in Appendix B.



The electron-boson coupling Hamiltonian $\hat\cH^{\rm e-b}_{\lm}$ (used for both photons, $\lambda = \ell$ for ``light,'' and phonons, $\lambda = s$ for ``sound''), reads:
\Eq{
\hat\cH^{\rm e-b}_{\lm}=\sum_{k,\bq}  \hat {\tb c}_k\dg U_\lm (\bq) \hat {\tb c}_{k+q} \hat b_{\lm,\bq}\dg +\rm {h.c.}
\label{eq:HamElBos}
}
Here $\hat b_{\lm,\bq}$ annihilates a photon (for $\lm=\ell$) or an acoustic phonon (for $\lm=\ys$), with crystal momentum $\bq=\bR{q,\bq_\perp}$, and frequency $\w_\lm(\bq)=v_\lm |\bq|$, where $v_\lm$ is the speed of light or sound, respectively.
The first component of $\vec{q}$, denoted $q$, is the crystal momentum component parallel to the wire, and $\vec{q}_\perp$ represents its orthogonal component(s).
(For photons, $\vec{q}_\perp$ has two components, while for phonons, $\vec{q}_\perp$ has one or two components, depending on whether $d = 2$ or $d = 3$.)
The microscopic details of the electron-photon and electron-phonon couplings are captured by the functions $U_{\lambda}(\vec{q})$.

We take the coupling between electrons and acoustic phonons polarized along the wire to be \cite{Bockelmann1990}
\Eq{
U_\ys(\bq,\w)=g_\ys P(q)q \bR{\frac{a v_\ys}{\w}}^\half.
\label{eq:ElPhInt}
}
Here $g_\ys$ is a coupling parameter, and we take the orbital coupling matrix $P(q)$ to be identity for small $q$.

For the electron-photon coupling we take the simple ($\vec{q}$-independent) form $U_\ell=g_\ell \s^x$, where $g_\ell$ is a coupling parameter and $\s^x$ is an orbital-space Pauli matrix [see Eq.~\eqref{eq:Hamiltonian}]. More realistic models for electron-photon coupling would not change the qualitative results of our analysis.

We work in the regime of weak system-bath coupling, where close to the steady state the electronic density matrix is well-described in terms of populations of the Floquet eigenstates~\cite{Seetharam2015,Esin2018}. 
These populations are given by $f_{k\n}(t)=\av{\hat\f_{k\n}\dg(t) \hat\f_{k\n}(t)}$, where $\hat\f\dg_{k\n}(t)$ creates an electron in the Floquet state $e^{-i\varepsilon_{k\nu}t}\ket{\f_{k\n}(t)}$ \footnote[100]{The coherences between $\n=-1$ and $\n=1$ can be neglected if $V \ta_{\rm scat}\gg\hb$, where $\ta_{\rm scat}$ is the typical scattering time-scale\cite{Seetharam2015}.}.

The system-bath coupling induces transitions between Floquet states. 
As a result, the populations $\{f_{k\nu}\}$ evolve according to the kinetic equation:
\EqS{
\dot f_{k\n}(t)=&\sum_{k'\n'\lm} [I_{\lm,\n\n'}(k,k')-I_{\lm,\n'\n}(k',k)],
\label{eq:Boltzmann}
}
where
\EqS{
I_{\lm,\n\n'}(k,k')=-\sum_l W^{(l)}_{\lm,\n\n'}(k,k')f_{k\n}(1-f_{k'\n'})
\label{eq:ScatteringPhonon}
}
describes the rate of electron transitions from state $k'$ in Floquet band $\n'$ to state $k$ in Floquet band $\n$.
For a zero temperature bath, the rate $W^{(l)}_{\lm,\n\n'}(k,k')$ in Eq.~(\ref{eq:ScatteringPhonon}) describes the corresponding scattering rate for a single electron in an otherwise empty system, involving spontaneous emission of a boson (phonon or photon) with energy $\hb\w_l=\ve_{k\n}-\ve_{k'\n'}+l\hb\W$.
From the Floquet Fermi's golden rule, this rate is given by:
\Eq{
W^{(l)}_{\lm,\n\n'}(k,k')=\frac{2\p}{\hb}\abs{U^{(l)}_{\lm,\n\n'}(k,k')}^2\ro_{\lm}(k-k',\w_l),
\label{eq:Intrinsicrate}
}
where $U^{(l)}_{\lm,\n\n'}(k,k')=\sum_m\braoket{\f^m_{k\n}}{U_{\lm}(k-k',\w_l)}{\f^{m-l}_{k'\n'}}$ is the matrix element associated with electron-phonon or electron-photon coupling.
Here $\ro_{\lm}(q,\w)$ denotes the density of states for $\lambda$-type boson emission with fixed momentum transfer $q$ along the direction of the nanowire. 

For small momentum transfer, $|q|\ll \w/v_\lm$, the densities of states for photon and phonon reservoirs read $\ro_\ell(q,\w)=\ro^0_\ell\cdot (a\w/v_\ell)\Q(\w-v_\ell|q|)$, and $\ro_\ys(q,\w)=\ro^0_\ys\cdot (a\w/v_\ys)^\h\Q(\w-v_\ys|q|)\Q(\w_D-\w)$, respectively. For acoustic phonons in a homogeneous crystal in $d$-dimensions, $\h=d-2$. Here $\ro^0_\ys$ and $\ro^0_\ell$ are constants in $q$ and $\w$, and $\ro_\ys(q,\w)$ is cut off at Debye frequency, $\w_D$, which we assume to be in the range $V<\hb\w_D<\D$, where $\D$ is the gap at the quasienergy zone edge $\ve=\hb\W/2$, see Fig.~\ref{fig:PhaseDiagram}a. The condition $\hbar \omega_D < \Delta$ ensures the absence of phonon-mediated Floquet-Umklapp processes, corresponding to transitions with $l>0$ in Eq.~\eqref{eq:Intrinsicrate} \cite{Seetharam2015}.

\section{Floquet metal to insulator phase transition}
In the steady-state, the majority of excited electrons ``pile up'' in the two ``valleys'' in the upper Floquet band near the resonance points, cf. Fig.~\ref{fig:PhaseDiagram}a, giving rise to the ``bottleneck'' effect \cite{Seetharam2015}. This effect is due to the larger rates associated with scattering processes ``incoming'' into the minima near the resonance points, compared with ``outgoing'' ones. The  incoming processes are mostly due to intraband relaxation of electrons occupying high quasienergies in the upper Floquet band, which scatter to states near the bottom of the upper Floquet band. The outgoing processes are interband relaxation processes, in which electrons near the bottom of the upper Floquet band scatter to states near the top of the lower Floquet band. The incoming rates are dominant due to the larger phase space of target states with large electron-phonon matrix elements in the case of intraband transitions. Due to particle-hole symmetry in our model \footnote[3]{We expect our qualitative results to hold also in systems with no particle-hole symmetry.}, the holes form a mirror imaged population in the lower Floquet band.

We expect the intraband relaxation rates, in which electrons occupying high quasienergies in the upper Floquet band scatter to states near the bottom of the upper Floquet band, to be dominant over the rates of interband relaxation processes, in which electrons near the bottom of the upper Floquet band scatter to states near the top of the lower Floquet band.
The reason for the difference between the rates is essentially due to the larger phase space of target states with large electron-phonon matrix elements in the case of intraband transitions.
As a result, in steady-state the majority of excited electrons ``pile up'' in the two ``valleys'' in the upper Floquet band near the resonance points, cf. Fig.~\ref{fig:PhaseDiagram}a, giving rise to the ``bottleneck'' effect \cite{Seetharam2015}. Due to particle-hole symmetry in our model \footnote[3]{We expect our qualitative results to hold also in systems with no particle-hole symmetry.}, the holes form a mirror imaged population in the lower Floquet band.

Within the regime of interest, the distributions of electrons in the bottom of the upper Floquet band ($f_{k+}$) and holes in the top of the lower Floquet band ($f_{k-}$) can to a good degree be approximately described by separate Fermi-Dirac distributions
\footnote[4]{Slightly better fit could be made by introducing a small momentum-dependent shift to the chemical potential. Also, we note that although the majority of excitations occupy the band-minima according to, a small but finite density of excitations, occupying high quasienergy levels is necessary to maintain this distribution.} related by $f_k\eqv f_{k+}=1- f_{k-}$, where
\Eq{
f_{k+}=\bS{1+e^{\bR{\ve_{k+}-\mef_+}/k_B\Tef}}\inv.
\label{eq:DistributionFunction}
}
This assertion will be verified by our numerical simulations, below.
Here $\mef_+$ and $\Tef$ are the effective chemical potential and temperature of the electrons in the upper Floquet band.
It follows that the effective chemical potential for holes is $\mef_-=-\mef_+$, and their effective temperature is equal to the temperature of the electrons, $\Tef$. For convenience, we define a chemical potential for electrons, counted from the bottom of the band, $\mef\eqv\mef_+-\ve_{k_R+}$.

\begin{figure}
  \centering
  \includegraphics[width=8.6cm]{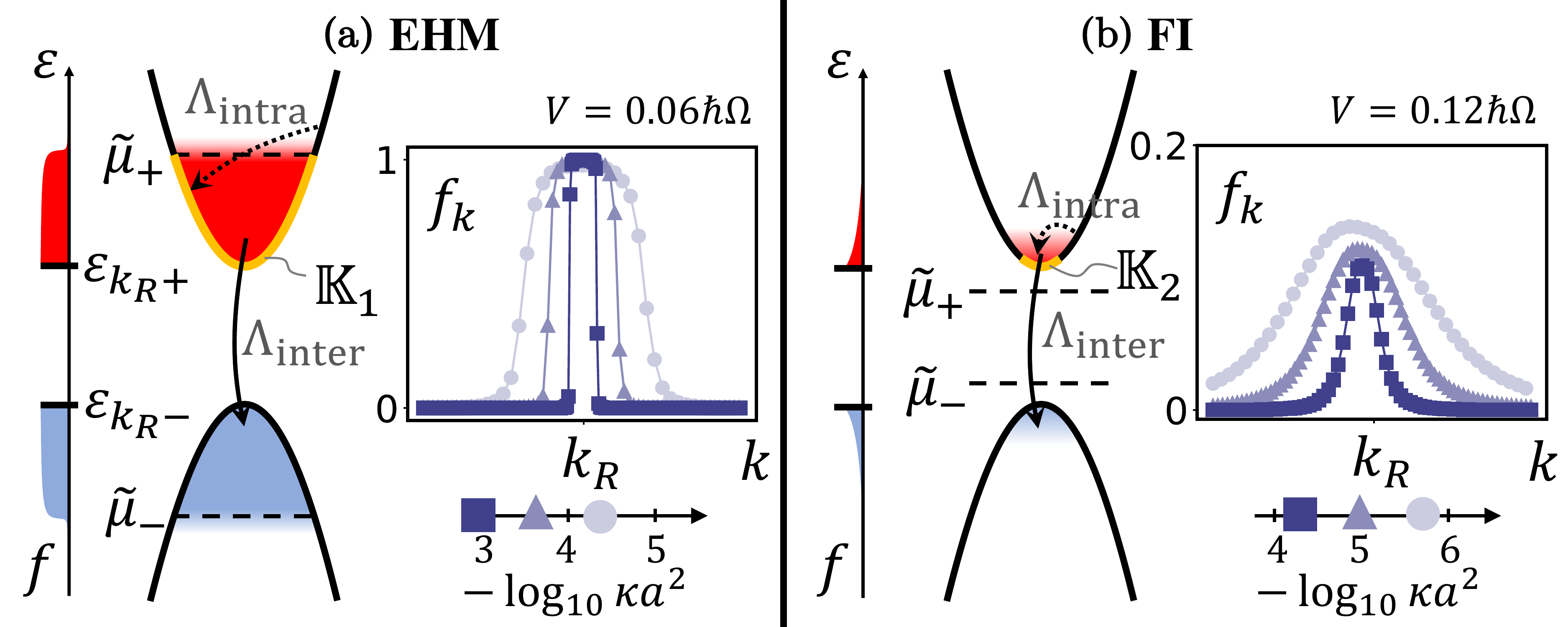}\\
  \caption{The bottleneck effect. Relaxation processes leading to the steady state near the minimum and maximum of the Floquet bands. (a) The EHM phase - effective chemical potentials for electrons ($\mef_+$) and holes ($\mef_-$) are inside the corresponding bands. Intra-band relaxation (dotted arrow) predominantly occurs across the chemical potential, connecting states within the range of $\Tef$ around the chemical potential. The orange colored part of the energy band, indicates the states hosting the density $n_\m$. Inset to (a): a steady-state distribution near $k=k_R$ in the EHM phase, for three values of the balance parameter, $\ka$, indicated on the logarithmic scale below the inset. Reducing the ``bottleneck'' parameter, $\a$, causes the chemical potential, $\mef_+$, to move toward the gap, until it crosses the band minima, at the critical value $\h$, causing the transition towards the FI phase. (b) FI phase - effective chemical potentials for the electrons and holes are in the Floquet gap. The orange colored part of the energy band, indicates the states hosting the density $n_T$. Intra-band relaxation processes (dotted arrow) are dominant in the minima of the band. Inset to (b): a steady-state distribution near $k=k_R$ in the FI phase.
  \label{fig:Processes}}
\end{figure}

In what follows, we will analyze the dependence of $\mef$ and $\Tef$ on the parameters of the system and the heat-baths.
To this end, we first develop a phenomenological model that captures the phase structure of the system and allows us to extract these dependencies. We will then corroborate these predictions with numerical simulations of the full kinetic equation [Eq.~\eqref{eq:Boltzmann}]. In the main text our analysis is done for zero bath temperature. The effects of finite bath temperature are discussed in Appendix A.

To build the phenomenological model, we seek two balance equations that must be satisfied by the populations of the Floquet bands in the steady state. The first balance equation fixes the value of the total excitation density, $n\eqv n_+=n_-=\int_{-\p/a}^{\p/a} \frac{dk}{2\p} f_{k}$, from the balance between inter-band excitation and relaxation processes \cite{Glazman1981,Glazman1983}. A given value of $n$ corresponds to many different combinations of $\mef$ and $\Tef$. A second equation, following from the balance of intra- and inter-band relaxation processes, provides the relation between $\mef$ and $\Tef$.

We now discuss the processes leading to the steady-state. Electron-hole excitations are predominantly generated by photon-mediated Floquet-Umklapp processes. In the low-excitation regime, $n a \ll 1$, which we consider throughout, photon-mediated processes excite electrons from an almost full to an almost empty Floquet band. Therefore, the excitation rate $\dot n|_{\rm rec}=\G_{\rm rec}$ is approximately independent of the steady-state distribution. Once excited, electron excitations quickly relax to the ``valleys'' of the upper Floquet band near $k=\pm k_R$ and accumulate there. A mirror imaged process applies for the holes.

Recombination of Floquet-electron-hole pairs occurs at longer timescales than the relaxation to the band minima and are primarily mediated by phonons. Their rate is proportional to the density of electron excitations in the upper Floquet band and hole excitations in the lower Floquet band i.e., $\dot n|_{\rm inter}=-\Lm_{\rm inter}n^2$ \cite{Seetharam2015}. The rate equation for the density of excited electrons due to inter- and intra- Floquet band processes then reads
\Eq{
\dot n=\G_{\rm rec}-\Lm_{\rm inter}n^2.
\label{eq:ContinuityForN}
}
The steady state solution is obtained upon setting $\dot n=0$, which yields
\Eq{
n=\ka^\half; \quad \ka\eqv\frac{\G_{\rm rec}}{\Lm_{\rm inter}}.
\label{eq:DefinitionKappa}
}
Here $\ka$ is the ``balance parameter'', denoting the balance between photon-mediated heating and phonon-mediated cooling processes. When $\ka=0$, the steady state resembles a zero-temperature Gibbs distribution for the Floquet quasienergies \cite{Galitskii1970}.

Next, we discuss the balance between the intra-band and inter-band relaxation processes. We begin by considering the situation deep in the EHM metal phase, where the excitations on top of the Floquet insulator state exhibit sharp Fermi surfaces, $\mef \gg k_B \Tef$. This regime is realized when interband relaxation is the rate limiting step in the relaxation of excited Floquet electron hole pairs. In order to determine the balance equation in this situation, we will partition the density of excited electrons, $n$, into two contributions: $n_\m$ and $\dl n$, corresponding to the total density of electrons with quasienergies between $\ve_{k_R}$ and $\mef_+$, and all other excited electrons, see Fig.~\ref{fig:Processes}a. Thus we define $n_\m=2\int_{\dK_1}\frac{dk}{2\p} f_k$, where $\dK_1=\bC{k>0|\ekr\le \ve_{k+}<\mef_+}$, and $\dl n=n-n_\m$. The 2 in the definition of $n_\m$ accounts for the contributions of the two valleys.

The dominant source rate for $n_\mu$ arises from the scattering of electrons with quasienergies above $\mef_+$ to empty states below $\mef_+$. Therefore, we estimate this rate by $\dot n_\m|_{\rm intra}=\Lm_{\rm intra} \dl h \dl n$, where $\Lm_{\rm intra}$ is the average intrinsic rate of collisions across $\mef_+$ and $\dl h\eqv\int_{\dK_1}\frac{dk}{2\p}(1-f_k)$. Processes contributing to $\Lm_{\rm intra}$ predominantly occur in a quasienergy window of width $k_B \Tef$ around the Fermi level, where the densities $\dl n$ and $\dl h$ are mostly concentrated. Therefore, we estimate the energy of emitted phonons by $\hb\w\sim k_B\Tef$. The density of states for such phonons is non-vanishing for momentum transfers $\hb|\dl q|<k_B\Tef/v_\ys$. Therefore, large momentum intra-band processes connecting populations near $k_R$ and -$k_R$ are forbidden for low temperature steady states, $k_B\Tef<2k_R\hb v_\ys$.

The main contribution to $\Lm_{\rm intra}$ comes from the largest momentum transfers allowed within each valley, $2k_F= \pi n$ (see Fig.~\ref{fig:Processes}a), due to the momentum dependence of $U_\ys(q)$ [Eq.~\eqref{eq:ElPhInt}]. Here $v_R$ is the velocity of the non-driven band at the resonance momentum. Since $\dl q$ is small on the scale of the Brillouin zone size, we evaluate the matrix element for phonon scattering by $|\sum_m\braoket{\f^m_{k_R,\n}}{P(\dl q)}{\f^m_{k_R+\dl q,\n}}|^2=1+\cO(\dl q)$, see text below Eq.\eqref{eq:ElPhInt}. Using Eq.~\eqref{eq:Intrinsicrate} with the density of states of phonons and Eq.~\eqref{eq:ElPhInt}, we estimate
\Eq{
\Lm_{\rm intra}\eqa a W_0 \bR{\frac{\mef}{V}}\bR{\frac{ak_B \Tef}{\hb v_\ys}}^{\h-1},
\label{eq:EstimateLintra}
}
with a constant $W_0$, independent of $\mef$ and $\Tef$.

The source rates for $n_\m$ are balanced by recombination of electrons in $n_\m$ with holes in the lower Floquet band. The rate of these outgoing interband processes is proportional to the density of electrons, $n_\m$, and the total density of holes, which is given by $n$. We thus estimate $\dot n_\m|_{\rm inter}=-\Lm_{\rm inter}n_\m n$. The average interband rate has two main contributions. One arises from ``vertical'' processes (shown by a black arrow in Fig.~\ref{fig:PhaseDiagram}a) with a momentum transfer $\dl q$. The second contribution is due to ``diagonal'' processes (shown by a purple arrow in Fig.~\ref{fig:PhaseDiagram}a) with a momentum transfer $\sim 2k_R+\dl q$.

We define the rate for interband scattering from the state $\ket{\f_{k_R,+}^0}$ to either $\ket{\f_{k_R+\dl q,-}^0}$ or $\ket{\f_{-k_R+\dl q,-}^0}$ for a fixed $\dl q$ as $W_{\rm inter}(\dl q)=W^{(0)}_{\ys,+-}(k_R,k_R+\dl q)+W^{(0)}_{\ys,+-}(k_R,-k_R+\dl q)$. In $W_{\rm inter}(\dl q)$, the first term corresponds to ``vertical'' processes, while the second to ``diagonal'' processes. The dominant term contributing to the total interband rate corresponds to $\dl q\eqa k_F$. For such a small momentum (recall that $k_F$ is the Fermi wave number corresponding to the small density of excited electrons/holes in each valley), we expand $W_{\rm inter}(\dl q)=\tilde W(\hb v_R\dl q/V)^{2\a}+O[(\hb v_R\dl q/V)^{2\a+1}]$, where $\tilde W$ is constant and
\Eq{
\a=\half\frac{\dpa \log W_{\rm inter}(\dl q)}{\dpa\log \dl q}\at{\dl q=0}.
\label{eq:AlphaDef}
}
In what follows, we refer to $\a$ as the ``bottleneck'' parameter, as it controls the relative strengths of the intra- and inter-band processes. Its value depends on the matrix element of $U_\ys$ [cf.~Eq.~\eqref{eq:ElPhInt}] between the Floquet states involved in the scattering process. The energy transfer of each process is approximately equal to the Floquet gap, $|V|$. Using Eq.~\eqref{eq:Intrinsicrate} and the phonon density of states, we estimate
\Eq{
\Lm_{\rm inter}\eqa a W_0 \bR{\frac{\mef}{V}}^\a\bR{\frac{aV}{\hb v_\ys}}^{\h-1}.
\label{eq:EstimateLinter}
}

\begin{figure}
  \centering
  \includegraphics[width=8.6cm]{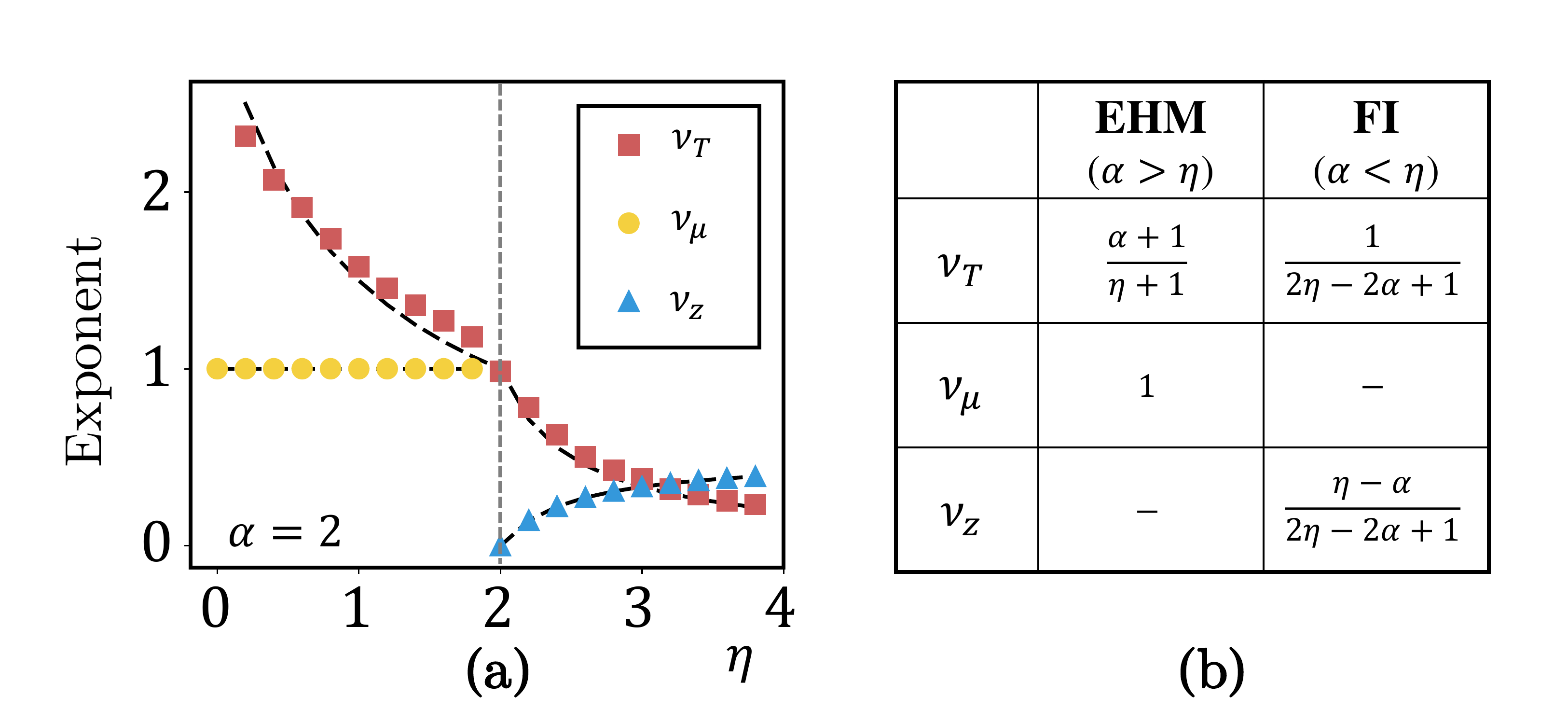}\\
  \caption{(a) Fitting parameters of the power laws of effective temperature, $\Tef/V\sim (\ka/\ka_0)^{\n_T}$, effective chemical potential, $\mef/V\sim (\ka/\ka_0)^{\n_\m}$, and the fugacity, $\zef\sim(\ka/\ka_0)^{\n_z}$, as a function of the exponent which appears in the $\w$-dependence in the density of states of phonons, for fixed $\a=2$. Black dashed lines correspond to an analytic prediction for the exponents from the phenomenological model (Eqs.~\eqref{eq:ContinuityForN} to \eqref{eq:MuTFI}), whose values are summarized in panel (b). Gray dashed line denotes the EHM-to-FI phase transition point at $\h=2$. The non-analytic behaviour of the exponents at $\a=\h$ implies the existence of a quantum critical point at this point when $\ka\to 0$, see Fig.~\ref{fig:PhaseDiagram}b. \label{fig:Exponenets}}
\end{figure}

Next, we combine all of the ingoing and outgoing rates for $n_\m$ that we found above, to arrive at the rate equation:
\Eq{
\dot n_\m=\Lm_{\rm intra} \dl h \dl n-\Lm_{\rm inter}n_\m n.
\label{eq:RateMuT}
}
We use Eq.~\eqref{eq:RateMuT} to obtain a relation between $\mef$ and $\Tef$ in the steady-state (when $\dot n_\m=0$). To this end, we express $\Lm_{\rm inter}$, $\Lm_{\rm intra}$ by their estimates as functions of $\mef$ and $\Tef$ [Eqs.~\eqref{eq:EstimateLintra} and \eqref{eq:EstimateLinter}]. We further approximate $n\eqa n_\m\eqa 4\mef D(\mef_+)$ and $\dl n\eqa \dl h\eqa 2k_B \Tef D(\mef_+)$, where $D(\ve)$ is the density of states near the parabolic minimum of the upper Floquet band. Using $D(\ve)=\bR{2\p\hb v_R}\inv [V/(\ve-\ekr)]^{\half}$, we obtain
\Eq{
\mef/k_B\Tef=c \bR{k_B \Tef/V}^{\frac{\h-\a}{\a+1}},
\label{eq:MuTEHM}
}
where $c$ is a numerical constant of the order of unity. Eq.~\eqref{eq:MuTEHM} is consistent with an EHM phase, in which $\mef\gg k_B\Tef$ and $k_B\Tef\ll V$ (yielding a sharp fermi surface), when $\a>\h$. Combining the expression for $n$ with Eqs.~\eqref{eq:DefinitionKappa} and \eqref{eq:MuTEHM}, we express $\mef$ and $\Tef$ as two separate functions of the balance parameter, $\ka$, yielding $\Tef/V\sim (\ka/\ka_0)^{\n_T}$, and $\mef/V\sim (\ka/\ka_0)^{\n_\m}$ where $\ka_0=(V/2\p\hb v_R)^2$. Explicit expressions for $\n_T$ and $\n_\m$ in terms of $\a$ and $\h$ appear in the table in Fig.~\ref{fig:Exponenets}b.


For $\a<\h$, the EHM is not a consistent steady state of the rate equations \eqref{eq:ContinuityForN} and \eqref{eq:RateMuT}. We will now therefore analyze the rate equation starting from the opposite limit, assuming a FI phase. In this phase, the chemical potential lies in the Floquet gap, $\mef<0$. Therefore, we approximate $f_k\eqa \zef e^{-\ve_{k+}/k_B \Tef}$, where $\zef=e^{-|\mef|/k_B\Tef}$ is the fugacity. The total density of excitations then reads $n\eqa 2\zef D(k_B\Tef)k_B \Tef$. Since the density $n_\m$ is not well defined here, we define the density $n_T=2\int_{\dK_2}\frac{dk}{2\p}f_k$. The range of integration $k\in \dK_2$ is over a small region of a length $\D k\ll D(k_B\Tef)k_B\Tef$ in momentum space, such that $n_T\eqa 2\zef \D k$ \footnote[44]{Other choices of $\D k$ will lead to the same power-law dependence.} (see Fig.~\ref{fig:Processes}b). This ensures $n_T\ll n$.
In addition, we define $\dl n'=n-n_T\eqa n$, and $\dl h'=2\int_{\dK_2}\frac{d k}{2\p}(1-f_k)\eqa 2\D k$. The rate equation for $n_T$ is similar to Eq.~\eqref{eq:RateMuT} upon replacing $n_\m$ with $n_T$ and $\dl n$, $\dl h$ with $\dl n'$, $\dl h'$. The electron and hole excitations in the FI phase spread over a quasienergy window of the order of $k_B\Tef$, in contrast to the range of order $\mef$ in the EHM phase. Therefore, we replace the $\mef$-dependence of $\Lm_{\rm intra}$ and $\Lm_{\rm inter}$ in Eqs.~\eqref{eq:EstimateLintra} and \eqref{eq:EstimateLinter} with $k_B\Tef$. Solving the resulting equation in the steady state, $\dot n_T=0$, we arrive at
\Eq{
\zef=c'(k_B\Tef/V)^{\h-\a},
\label{eq:MuTFI}
}
where $c'$ is a numerical coefficient of order unity, generically different from $c$ in Eq.~\eqref{eq:MuTEHM}. Eq.~\eqref{eq:MuTFI} yields a dependence of $\zef$ and $\Tef$ on $\ka$ for the FI phase, in the form of the power laws with exponents $\n_T$ and $\n_z$. The exponents are summarized in the table in Fig.~\ref{fig:Exponenets}b.

Note the difference between the exponents $\n_T$ in the FI and EHM phases. This difference, together with, the power laws for $\zef$ in the FI phase and $\mef$ in the EHM phase, indicate that the dependence on $\ka$ of these important quantities is discontinuous across the EHM-FI transition. This implies the existence of a quantum critical point at $\a=\h$ when $\ka\to 0$. Increasing $\ka$ increases the effective temperature of the steady state, $\Tef$, giving rise to a finite effective-temperature crossover above the critical point in the $\ka$-$\a$ plane, see Fig.~\ref{fig:PhaseDiagram}b.

To support the analysis above we numerically solve for the steady state of the full Floquet kinetic equation [Eq.~\eqref{eq:Boltzmann}] on a lattice of 5000 $k$-points, at half filling.
We fit the steady state to a Fermi-Dirac distribution of the electrons in the upper Floquet band to extract $\Tef$, $\mef$ and $\zef$. We then extract the power law scalings of the these three quantities as functions of $\ka$, see Fig.~\ref{fig:Exponenets}a. In these simulations we fix $\a=2$ by setting $A=0$ in Eq.~\eqref{eq:Hamiltonian} \footnote[55]{Note that for $A=0$, the contribution to the rate coming from both the ``vertical'' and ``diagonal'' processes (corresponding to momentum transfers $\sim 2k_F$ and $\sim 2k_R$ respectively) yield $\alpha=2$ in Eq.~\eqref{eq:AlphaDef}.}, and sweep the value of $\h$ from $\h=0$ to $\h=4$, across the critical value at $\h=2$. Fig.~\ref{fig:Exponenets}a shows the exponents $\n_T$, $\n_\m$, and $\n_z$ extracted numerically from the fit to power-laws. We find a good agreement between our analytical and numerical results.

\begin{figure}
  \centering
  \includegraphics[width=8.6cm]{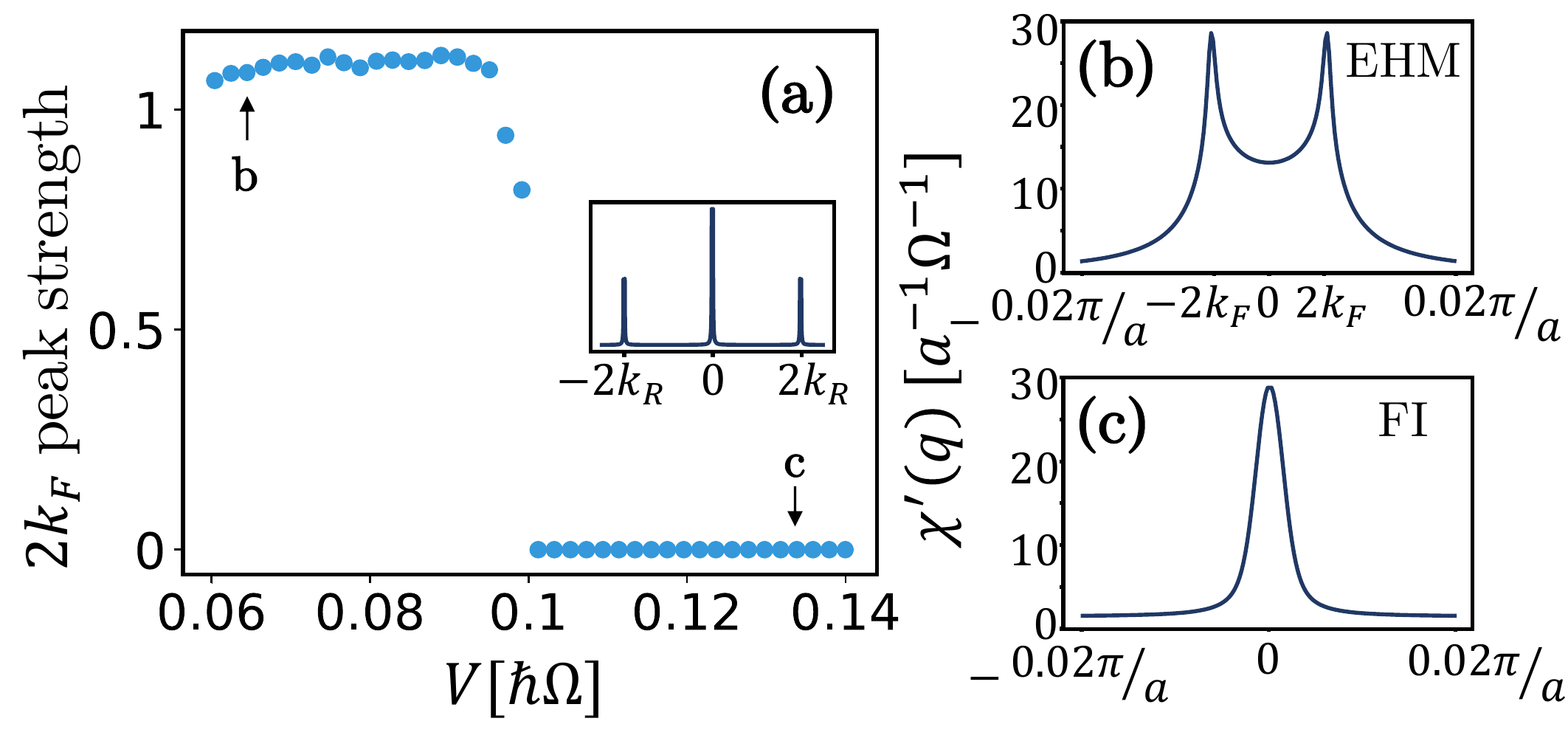}\\
  \caption{Response of the steady-state to density fluctuations. (a) The strength of the $2k_F$-peak, $\cF$, as function of the driving field strength, $V$, across the critical value $V_c=0.1\hb\W$. Inset: a full Fourier spectrum of the real part of the response function at zero frequency, $\ch'(q)$. (b), (c). Zoom in on the Fourier peak of $\ch'(q)$ near $q=0$. In the EHM phase (b), the Fourier transform exhibits $2k_F$ peaks, whereas in the FI phase (c), the function is smooth.
 \label{fig:Friedel}}
\end{figure}

\section{Experimental realization and signatures }
In this section we discuss how to experimentally induce and observe the transition between the EHM and the FI phase. The transition can be tuned by increasing the amplitude of the driving field. The EHM phase requires the driving amplitude to be below a critical value. To see why, note that the ``diagonal'' processes with large momentum transfer of $2k_R$ are only active when the Floquet gap is above a critical value $V_c=2 \hb v_\ys k_R$. (The Floquet gap approximately sets the energy of the phonons involved, which must be larger than the minimal phonon energy $2\hbar \omega \approx 2 \hb v_\ys k_R$.) Therefore, when $V<V_c$, only ``vertical'' processes corresponding to $\a=2$ contribute, while for $V>V_c$, the ``diagonal'' processes become dominant, yielding $\a=0$ \footnote[67]{The value of $\a$ depends on the overlap of the eigenfunctions. Therefore, it may differ if the system has extra symmetries.}. For $\h=1$, corresponding to phonons in three-dimensions, an EHM is obtained for $V<V_c$, while the FI is stabilized for $V>V_c$.

 The difference between the phases is manifested for example in the response of the system to density perturbations. To this end, we compute the density-density response function averaged over the driving period, $\ch(x,t)=\frac{\Q(t)}{i\hb\cT}\int_0^\cT d\ta\tr{\hat \vr_{\rm st}(t)\com{\hat n(x,t+\ta)}{\hat n(0,\ta)}}$, where $\hat\vr_{\rm st}$ is the steady state density matrix and $\hat n(x)=\hat{\tb c}_x\dg\hat{\tb c}_x$ is the density operator, with $\hat {\tb c}_x=\int dk e^{ikx}\hat{ \tb c}_k$. Using $\hat\vr_{\rm st}(t)=\prod_{k\n}[f_{k\n}\hat \f\dg_{k\n}(t) \hat \f_{k\n}(t)+(1-f_{k\n})\hat \f_{k\n}(t) \hat \f\dg_{k\n}(t)]$ we obtain the Floquet-Lindhard function \cite{Torres2005a}
\Eq{
\ch(q,\w)=\sum_{\substack{\n\n'l}}\int \frac{dk}{2\p}\frac{\cM_{\n\n'}^{(l)}(k,k-q)(f_{k-q\n'}-f_{k\n})}{\hb\w+\ve_{k-q\n'}-\ve_{k\n}-l\hb\W+i0^+},
\label{eq:Lindhard}
}
where $\cM_{\n\n'}^{(l)}(k,k')=|\sum_m \braket{\f_{k\n}^m}{\f_{k'\n'}^{m-l}}|^2$.

We numerically compute the Floquet-Lindhard function, $\ch(q,\w)$, as a function of driving amplitude across the EHM-to-FI transition \footnote[34]{To resolve the transition numerically, in these simulations we increased $v_\ys$ in the phonon density of states for transitions involving large phonon momenta $|q|>k_R$, see Appendix C for details.}. The inset to Fig.~\ref{fig:Friedel}a shows the real part of the response function $\ch'(q,0)=\text{Re}[\ch(q,0)]$. In the FI phase, $\ch'(q)$ exhibits large peaks at zero wavenumber, and at wavenumbers connecting the two valleys at $k_R$ and $-k_R$. In addition, due to a sharp Fermi surface in the EHM phase, each peak splits into two peaks separated by $2k_F$.
In particular, $\ch'(q)$ exhibits two peaks at $q=\pm 2k_F$. The splitting of these peaks is absent in the FI phase (see Figs.~\ref{fig:Friedel}b and \ref{fig:Friedel}c). Experimental signatures of the split peaks are Friedel oscillations in the screening potential. We draw the strength of the $2k_F$ peaks, defined as $\cF=\frac{\ch'(2k_F)-\ch'(0)}{\ch'(0)}$, as a function of the driving field strength, $V$ (along the red arrow in Fig.~\ref{fig:PhaseDiagram}b), see Fig.~\ref{fig:Friedel}a. The peaks disappear for $V>V_c$, for which the steady state is in the FI phase.

\section{Discussion}
To appreciate the physical scales, we estimate the effective temperature and chemical potential in periodically driven semiconductors in the EHM phase. We evaluate the rates of radiative recombination, and phonon-mediated relaxation associated with the hot-electron lifetime by $\ta_{\rm  rr}\eqa 1 \text{ ns}$ and $\ta_{\rm he}\eqa 10 \text{ fs}$, respectively \cite{Sundaram2002}. This yields the estimate $\ka a^2\eqa \frac{k_R a}{\p}\frac{\ta_{\rm he}}{\ta_{rr}}= 1.5\times 10^{-6}$. We take typical carrier and sound velocities $v_R= 1\times 10^5 \text { m/sec}$ and $v_\ys= 1\times 10^3 \text { m/sec}$, and $V=1.9 \text{ meV}$, which yield $\ka_0 a^2\eqa 8.4\times 10^{-6}$ and $\ka/\ka_0\eqa 0.18$. Therefore, the excitations in EHM phase coupled to a three-dimensional phonon reservoir yield an excitation density that corresponds to about $0.02\%$ of the Brillouin zone, with $\mef\eqa 0.34\text{ meV}$, and $\Tef\eqa1.65\text{ K}$, corresponding to $k_B\Tef/\mef\eqa0.4$.

In this work, we uncovered a transition between EHM and FI phases in a driven one-dimensional electronic system. Our results can be generalized
to other one- and two-dimensional resonantly driven Floquet-Bloch systems. In two-dimensional systems, we expect the system still supports EHM and FI phases that can be accessed via the drive strength. Studying the features of the transition in two-dimensional systems, and, e.g.,  possibilities for controlling their behavior by reservoir engineering, are interesting directions for future study.


\begin{acknowledgments}
\section{Acknowledgments}
We would like to thank Tobias Gulden, Gaurav K. Gupta, Vladimir Kalnitsky, Gali Matsman, and Ari Turner for illuminating discussions. David Cohen and Yan Katz for technical support.
N. L. acknowledges support from the European Research Council (ERC) under the European Union Horizon 2020 Research and Innovation Programme (Grant Agreement No. 639172), and from the Israeli Center of Research Excellence (I-CORE) ``Circle of Light''. M. R. gratefully acknowledges the support of the European Research Council (ERC) under the European Union Horizon 2020 Research and Innovation Programme (Grant Agreement No.678862), and the Villum Foundation. I. E. acknowledges support from the Ministry of Science and Technology, Israel.

\end{acknowledgments}

\begin{appendix}

\section{Appendix A: Effect of non-zero temperature heat-baths}

\begin{figure}
  \centering
  \includegraphics[width=8.6cm]{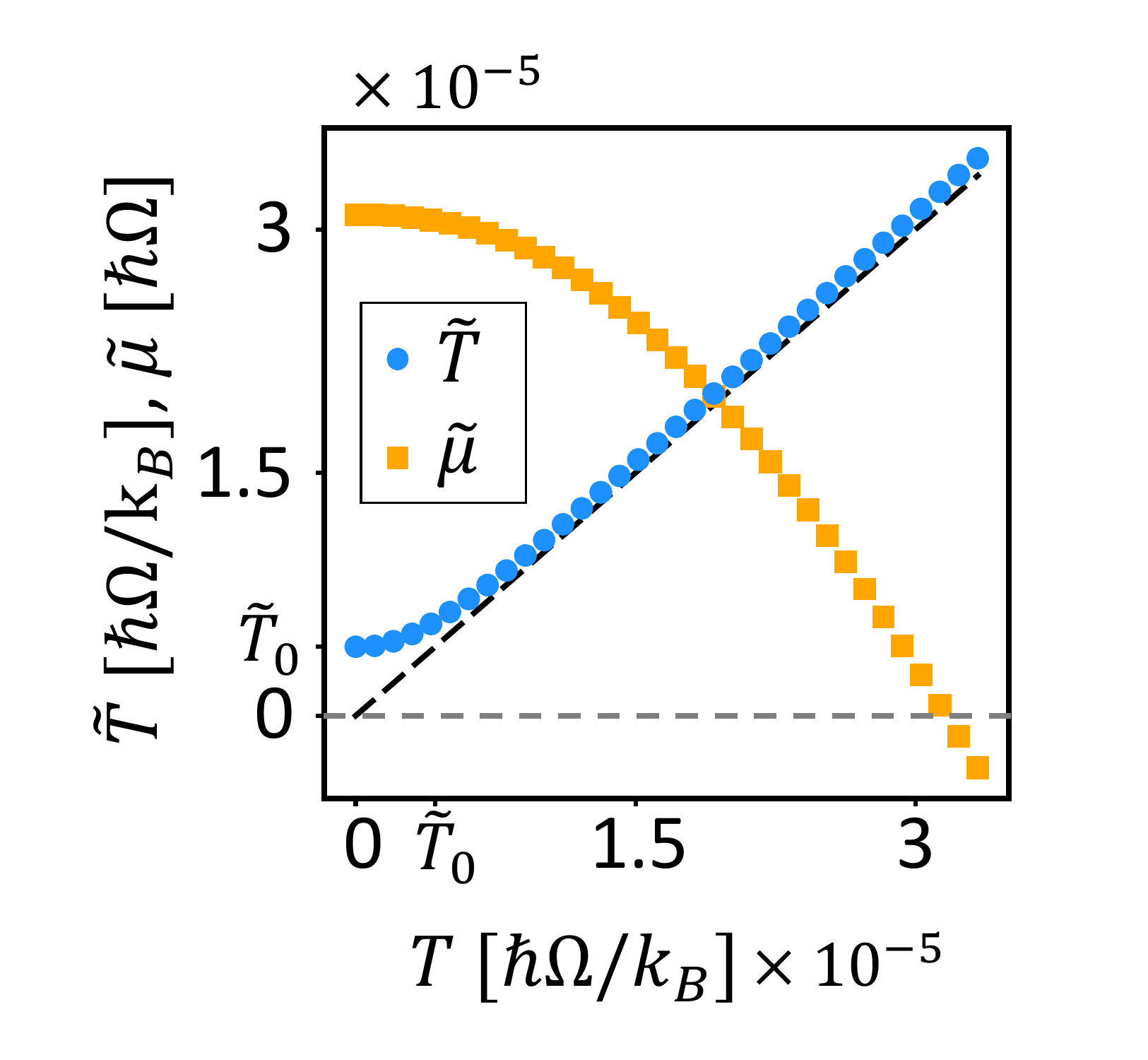}\\
  \caption{An effective temperature and chemical potential of the steady state in the EHM phase (for $V=0.06\hb\W$) as function of the heat-bath temperature, $T$. For $T\gtrsim \Tef_0$, the steady state thermalizes to $T$. The effective chemical potential first slowly changes with $T$, but as the sharp Fermi surface condition becomes less prominent, its slope grows until it becomes linear in $T$, as in the FI phase. \label{fig:Thermalization}}
\end{figure}

In this section, we analyze the case of non-zero temperature heat-baths. In this case, the density of excitations [in Eq.~\eqref{eq:ContinuityForN}] is balanced by additional thermal excitation and relaxation processes predominantly mediated by hot phonons. The effect of thermal processes is substantial when the temperature of the heat-baths, $T$, is larger than the ``intrinsic'' effective temperature of the steady state, $\Tef_0$. This refers to the effective temperature of the steady-state when the bath temperature is taken to be zero. In this situation, the thermal rates surpass the non-equilibrium rates and the steady-state in each Floquet band thermalizes to the heat-baths' temperature, see Fig.~\ref{fig:Thermalization}.

In an intermediate temperature regime, corresponding to $k_B T<V$, thermal fluctuations are not sufficient to induce transitions across the Floquet gap, and the electron and hole densities only slightly change due to the dependence of the inter-band relaxation rate on the temperature. Their corresponding chemical potentials, though, must be adjusted to maintain the new temperature and densities. This effect is more prominent in FI phase, where $\mef$ changes linearly with $T$ to leading order. In the EHM phase, the chemical potentials for electrons and holes are approximately constant as long as the sharp Fermi surface condition ($\mef\gg k_B\Tef$) is satisfied. For higher temperatures, the Fermi-surfaces are spoiled and the chemical potentials move linearly in $T$ toward the Floquet gap. For even higher temperatures above $V/k_B$ the two chemical potentials merge to a single one in the middle of the gap, leading to a thermalization of the distributions in the upper and lower Floquet bands to a single Gibbs-like distribution for the Floquet quasi-spectrum \cite{Galitskii1970}.

\section{Appendix B: Effects of electron-electron interactions}
\begin{figure}
  \centering
  \includegraphics[width=8.6cm]{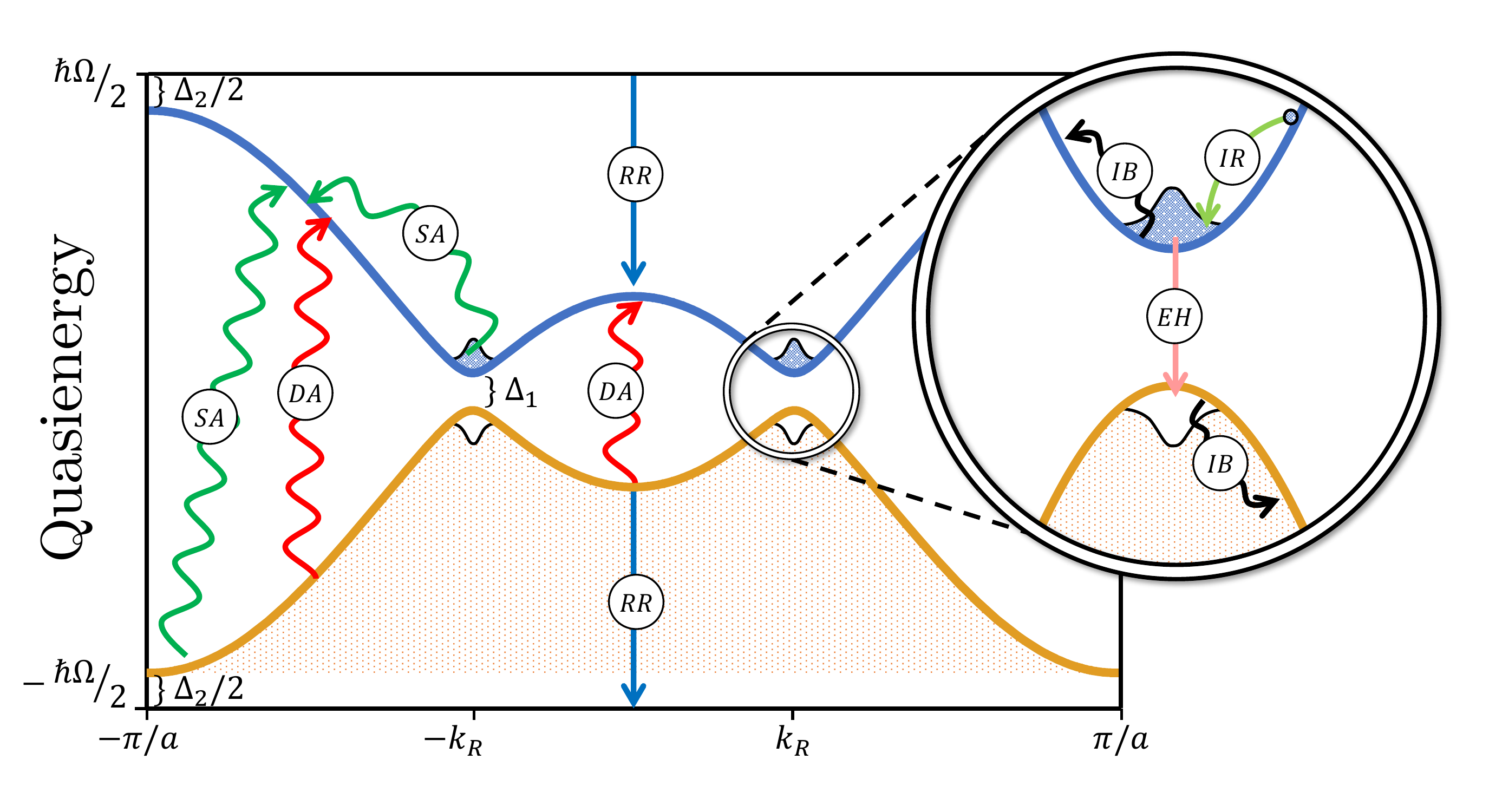}\\
  \caption{The Floquet quasi-spectrum of the Hamiltonian in Eq.~\eqref{eq:Hamiltonian} with a schematic drawing of the steady-state. The parameters of $H$ are $B=1/7\hb\W$, $A=2B$, and $\D_1=0.2A$, it follows that $\D_2=1/7\hb\W$. Electron and hole excitations occupy the minima and the maxima of the upper and lower Floquet bands, respectively. Solid arrows represent processes induced by phonon and photon emission: `RR' -- denotes radiative recombination processes; `IR', `EH' -- denote phonon mediated intra-band, and electron-hole recombination process, respectively. Wiggly arrows represent electron-electron collision processes: `IB' -- denotes intra-band interaction processes, fully conserving the quasienergy; `DA' and `SA' -- denote double and single Auger processes, respectively, conserving the quasienergy only modulo $\hb \W$. \label{fig:Interactions}}
\end{figure}

In this section we discuss the effects of electron-electron interactions on the steady state. We identify three main categories of interaction processes, indicated in Fig.~\ref{fig:Interactions} by wiggly arrows. Intra-band (IB) collisions refer to processes where the two electrons after the collision scatter into states in their original Floquet bands. The conservation of crystal momentum and quasienergy predominantly activates the collisions of electrons in the upper Floquet band with holes in the lower Floquet band near $k=\pm k_R$. Since the intra-band collisions do not change the densities of excitations, they primarily thermalize the distributions within each band. To demonstrate this effect, we assume  the steady-state distribution described by Eq.~\eqref{eq:DistributionFunction} with a small $k$-dependent correction to chemical potential $\mef\to\mef+\dl\m_k$. Then the net rate for scattering between electrons and holes, where the former particle scatters from $k$ to $k'$, reads
\EqS{
I_{\rm ee}(k,k')=&\sum_{q} W^{(0)}_{\rm ee}(k,k';q)f_{k}\bar f_{k'}\bar f_{-k+q} f_{-k'+q}(1-e^{\frac{\D\mef}{k_B\Tef}}).
\label{eq:InteractionRate}
}
Here $\D\mef=\dl\m_k-\dl\mef_{k'}-\dl\mef_{-k+q}+\dl\mef_{-k'+q}$, and $W^{(l)}_{\rm ee}(k,k';q)$ is nonzero only when $\ve_k-\ve_{k'}=\ve_{-k+q}-\ve_{-k'+q}+l\hb\W$. The scattering rate is linear in $\D \mef$ for small deviations from the Gibbs distribution.

Next, we consider the Double Auger (DA) and Single Auger (SA) processes.  The dominant effect of those on the steady-state distribution would be providing additional channels for excitations through non-radiative Auger recombination. To conserve the quasienergy, the SA and DA processes require an absorption of a photon from the driving field, hence their rate is suppressed by a factor of $(V/\hb\W)^2$. The DA processes correspond to collisions of two electrons in the lower Floquet band, which are kicked into two states in the upper Floquet band. Therefore the density of excitations is changed due to these processes by a rate that is approximately independent of the steady state, $\dot n|_{\rm DA}=2\G_{\rm DA}$. The SA  processes correspond to collisions of an electron in the lower Floquet band with another electron in the upper Floquet band near $k=\pm k_R$, scattering both to states in the upper Floquet band. The rate of such a process is proportional to the density of excitations, i.e., $\dot n|_{\rm SA}=\g_{\rm SA}n$, and hence can be neglected with respect to $\dot n|_{\rm DA}$ in the limit of small $n$. Therefore Auger processes modify Eq.~\eqref{eq:ContinuityForN}, leading to a renormalized steady-state excitation density, $n'$ and $\ka'$,
\Eq{
\ka'=\frac{2\G_{\rm DA}+\G_{\rm rec}}{\Lm_{\rm inter}}; \quad n'=\ka'^\half.
}

\section{Appendix C: Numerical simulations}
In order to observe the EHM-to-FI transition as a function of driving amplitude, our numerical simulations needed to satisfy two requirements. First, the number of $k$ points around $k_R$ must be large enough to resolve the momentum distribution to a degree which will allow us to differentiate between the EHM and the FI phase. This sets a requirement on the Floquet band gap, $V$, since the number of $k$ points in the parabolic region near $k_R$ is set by $a V/h v_R$, where $v_R$ is the velocity of the electrons at $k_R$ absent the drive. Second, the Floquet bandgap at the transition between the FI and the EHM phase is set by $2k_R\hb v_\ys$ where $v_\ys$ is the speed of sound.

To satisfy both these requirements, and to obtain a Floquet bandgap which is large enough to allow us to resolve the transition numerically, we artificially increased the speed of sound of acoustic phonons (relative to the electronic velocity $v_R$). This artificial increase of $v_\ys$ introduces kinematic constraints in electron-phonon collision processes within each valley \cite{Ziman1956}.

As we show below, for physically relevant parameters, and in particular, for physical values of $v_\ys$ in semiconductors,  such kinematic constraints are actually expected to be negligible in the steady state. Therefore, in order to prevent the artificial increase of $v_\ys$ from introducing kinematic constraints into our numerical simulations, we used a modified density of states for small momentum phonons. The modified density of states used in the simulation reads
\Eq{
\ro_\ys(q,\w)=\ro^0_\ys\cdot (a\w/v_\ys)^{\h}\Q(\w-\w_D)F(q,\w),
\label{eq:DOSapprox1}
}
where
\Eq{
F(q,\w)=\condf{\Q(\w) & ,|q|<q_T\\ \Q(\w-v_\ys |q|) & ,|q|\ge q_T}.
\label{eq:DOSapprox2}
}
For the definition of $\ro_\ys(q,\omega)$, see Eq.~\eqref{eq:Intrinsicrate}. The threshold momentum, $q_T$, is chosen to satisfy $2k_R\gg q_T\gg 2k_F$. With this choice for $q_T$, the total rate of scattering processes involving phonons with momenta close to $q_T$ (along the wire) are strongly suppressed in the steady state. This suppression is due to the small occupancy of electrons (holes) in quasi-momenta in the upper (lower) Floquet band which can scatter off phonons and transfer momentum $\pm q_T$ to the phonon bath.

We now show that for physically relevant parameters, kinematic constraints in electron-phonon collision processes within each valley are negligible in the steady state of both the FI and EHM phases. There are two main processes where the value of $v_\ys$ is important: in the FI phase, relaxation to the band bottom is kinematically constrained for states in the upper Floquet band below the energy $2mv_\ys^2$, where $m=\frac{V}{2v_R^2}$ is the effective mass at the band minima. Therefore, for $k_B\Tef\gg V (v_\ys/v_R)^2$, most of the relaxation rates are not affected by this constraint.
In the EHM phase, processes connecting opposite Fermi points in the same valley (near $\pm k_R$) are allowed when the energy transfer is larger than $2k_F \hb v_\ys$. Thus for $k_B\Tef\gg 2k_F \hb v_\ys$, the majority of such processes are unaffected by the constraint. Employing $\hb k_F=\sqrt{2m \mef}$, we arrive at the condition $(k_B\Tef)^2\gg 4 V\mef(v_\ys/v_R)^2$.
For most semiconductors, the inequality $v_\ys\ll v_R$ holds. In particular, for the physical parameter regime of interest discussed in the main text we evaluate $k_B\Tef/V=0.075$ and $\mef/V=0.18$ for $v_\ys/v_R=10^{-2}$, which meets the conditions for the validity of Eq.~\eqref{eq:DOSapprox1} and Eq.~\eqref{eq:DOSapprox2}

\end{appendix}


\begin{thebibliography}{59}%
\makeatletter
\providecommand \@ifxundefined [1]{%
 \@ifx{#1\undefined}
}%
\providecommand \@ifnum [1]{%
 \ifnum #1\expandafter \@firstoftwo
 \else \expandafter \@secondoftwo
 \fi
}%
\providecommand \@ifx [1]{%
 \ifx #1\expandafter \@firstoftwo
 \else \expandafter \@secondoftwo
 \fi
}%
\providecommand \natexlab [1]{#1}%
\providecommand \enquote  [1]{``#1''}%
\providecommand \bibnamefont  [1]{#1}%
\providecommand \bibfnamefont [1]{#1}%
\providecommand \citenamefont [1]{#1}%
\providecommand \href@noop [0]{\@secondoftwo}%
\providecommand \href [0]{\begingroup \@sanitize@url \@href}%
\providecommand \@href[1]{\@@startlink{#1}\@@href}%
\providecommand \@@href[1]{\endgroup#1\@@endlink}%
\providecommand \@sanitize@url [0]{\catcode `\\12\catcode `\$12\catcode
  `\&12\catcode `\#12\catcode `\^12\catcode `\_12\catcode `\%12\relax}%
\providecommand \@@startlink[1]{}%
\providecommand \@@endlink[0]{}%
\providecommand \url  [0]{\begingroup\@sanitize@url \@url }%
\providecommand \@url [1]{\endgroup\@href {#1}{\urlprefix }}%
\providecommand \urlprefix  [0]{URL }%
\providecommand \Eprint [0]{\href }%
\providecommand \doibase [0]{http://dx.doi.org/}%
\providecommand \selectlanguage [0]{\@gobble}%
\providecommand \bibinfo  [0]{\@secondoftwo}%
\providecommand \bibfield  [0]{\@secondoftwo}%
\providecommand \translation [1]{[#1]}%
\providecommand \BibitemOpen [0]{}%
\providecommand \bibitemStop [0]{}%
\providecommand \bibitemNoStop [0]{.\EOS\space}%
\providecommand \EOS [0]{\spacefactor3000\relax}%
\providecommand \BibitemShut  [1]{\csname bibitem#1\endcsname}%
\let\auto@bib@innerbib\@empty
\bibitem [{\citenamefont {Kitagawa}\ \emph {et~al.}(2010)\citenamefont
  {Kitagawa}, \citenamefont {Berg}, \citenamefont {Rudner},\ and\ \citenamefont
  {Demler}}]{Kitagawa2010}%
  \BibitemOpen
  \bibfield  {author} {\bibinfo {author} {\bibfnamefont {T.}~\bibnamefont
  {Kitagawa}}, \bibinfo {author} {\bibfnamefont {E.}~\bibnamefont {Berg}},
  \bibinfo {author} {\bibfnamefont {M.}~\bibnamefont {Rudner}}, \ and\ \bibinfo
  {author} {\bibfnamefont {E.}~\bibnamefont {Demler}},\ }\href {\doibase
  10.1103/PhysRevB.82.235114} {\bibfield  {journal} {\bibinfo  {journal}
  {Physical Review B}\ }\textbf {\bibinfo {volume} {82}},\ \bibinfo {pages}
  {235114} (\bibinfo {year} {2010})}\BibitemShut {NoStop}%
\bibitem [{\citenamefont {Wang}\ \emph {et~al.}(2013)\citenamefont {Wang},
  \citenamefont {Steinberg}, \citenamefont {Jarillo-Herrero},\ and\
  \citenamefont {Gedik}}]{Wang2013}%
  \BibitemOpen
  \bibfield  {author} {\bibinfo {author} {\bibfnamefont {Y.~H.}\ \bibnamefont
  {Wang}}, \bibinfo {author} {\bibfnamefont {H.}~\bibnamefont {Steinberg}},
  \bibinfo {author} {\bibfnamefont {P.}~\bibnamefont {Jarillo-Herrero}}, \ and\
  \bibinfo {author} {\bibfnamefont {N.}~\bibnamefont {Gedik}},\ }\href@noop {}
  {\bibfield  {journal} {\bibinfo  {journal} {Science}\ }\textbf {\bibinfo
  {volume} {342}},\ \bibinfo {pages} {453} (\bibinfo {year}
  {2013})}\BibitemShut {NoStop}%
\bibitem [{\citenamefont {Rudner}\ \emph {et~al.}(2013)\citenamefont {Rudner},
  \citenamefont {Lindner}, \citenamefont {Berg},\ and\ \citenamefont
  {Levin}}]{Rudner2013}%
  \BibitemOpen
  \bibfield  {author} {\bibinfo {author} {\bibfnamefont {M.~S.}\ \bibnamefont
  {Rudner}}, \bibinfo {author} {\bibfnamefont {N.~H.}\ \bibnamefont {Lindner}},
  \bibinfo {author} {\bibfnamefont {E.}~\bibnamefont {Berg}}, \ and\ \bibinfo
  {author} {\bibfnamefont {M.}~\bibnamefont {Levin}},\ }\href {\doibase
  10.1103/PhysRevX.3.031005} {\bibfield  {journal} {\bibinfo  {journal}
  {Physical Review X}\ }\textbf {\bibinfo {volume} {3}},\ \bibinfo {pages}
  {031005} (\bibinfo {year} {2013})}\BibitemShut {NoStop}%
\bibitem [{\citenamefont {Rechtsman}\ \emph {et~al.}(2013)\citenamefont
  {Rechtsman}, \citenamefont {Zeuner}, \citenamefont {Plotnik}, \citenamefont
  {Lumer}, \citenamefont {Podolsky}, \citenamefont {Dreisow}, \citenamefont
  {Nolte}, \citenamefont {Segev},\ and\ \citenamefont
  {Szameit}}]{Rechtsman2013}%
  \BibitemOpen
  \bibfield  {author} {\bibinfo {author} {\bibfnamefont {M.~C.}\ \bibnamefont
  {Rechtsman}}, \bibinfo {author} {\bibfnamefont {J.~M.}\ \bibnamefont
  {Zeuner}}, \bibinfo {author} {\bibfnamefont {Y.}~\bibnamefont {Plotnik}},
  \bibinfo {author} {\bibfnamefont {Y.}~\bibnamefont {Lumer}}, \bibinfo
  {author} {\bibfnamefont {D.}~\bibnamefont {Podolsky}}, \bibinfo {author}
  {\bibfnamefont {F.}~\bibnamefont {Dreisow}}, \bibinfo {author} {\bibfnamefont
  {S.}~\bibnamefont {Nolte}}, \bibinfo {author} {\bibfnamefont
  {M.}~\bibnamefont {Segev}}, \ and\ \bibinfo {author} {\bibfnamefont
  {A.}~\bibnamefont {Szameit}},\ }\href {\doibase 10.1038/nature12066}
  {\bibfield  {journal} {\bibinfo  {journal} {Nature}\ }\textbf {\bibinfo
  {volume} {496}},\ \bibinfo {pages} {196} (\bibinfo {year}
  {2013})}\BibitemShut {NoStop}%
\bibitem [{\citenamefont {Jotzu}\ \emph {et~al.}(2014)\citenamefont {Jotzu},
  \citenamefont {Messer}, \citenamefont {Desbuquois}, \citenamefont {Lebrat},
  \citenamefont {Uehlinger}, \citenamefont {Greif},\ and\ \citenamefont
  {Esslinger}}]{Jotzu2014}%
  \BibitemOpen
  \bibfield  {author} {\bibinfo {author} {\bibfnamefont {G.}~\bibnamefont
  {Jotzu}}, \bibinfo {author} {\bibfnamefont {M.}~\bibnamefont {Messer}},
  \bibinfo {author} {\bibfnamefont {R.}~\bibnamefont {Desbuquois}}, \bibinfo
  {author} {\bibfnamefont {M.}~\bibnamefont {Lebrat}}, \bibinfo {author}
  {\bibfnamefont {T.}~\bibnamefont {Uehlinger}}, \bibinfo {author}
  {\bibfnamefont {D.}~\bibnamefont {Greif}}, \ and\ \bibinfo {author}
  {\bibfnamefont {T.}~\bibnamefont {Esslinger}},\ }\href {\doibase
  10.1038/nature13915} {\bibfield  {journal} {\bibinfo  {journal} {Nature}\
  }\textbf {\bibinfo {volume} {515}},\ \bibinfo {pages} {237} (\bibinfo {year}
  {2014})}\BibitemShut {NoStop}%
\bibitem [{\citenamefont {Grushin}\ \emph {et~al.}(2014)\citenamefont
  {Grushin}, \citenamefont {G{\'{o}}mez-Le{\'{o}}n},\ and\ \citenamefont
  {Neupert}}]{Grushin2014}%
  \BibitemOpen
  \bibfield  {author} {\bibinfo {author} {\bibfnamefont {A.~G.}\ \bibnamefont
  {Grushin}}, \bibinfo {author} {\bibfnamefont {{\'{A}}.}~\bibnamefont
  {G{\'{o}}mez-Le{\'{o}}n}}, \ and\ \bibinfo {author} {\bibfnamefont
  {T.}~\bibnamefont {Neupert}},\ }\href {\doibase
  10.1103/PhysRevLett.112.156801} {\bibfield  {journal} {\bibinfo  {journal}
  {Physical Review Letters}\ }\textbf {\bibinfo {volume} {112}},\ \bibinfo
  {pages} {156801} (\bibinfo {year} {2014})}\BibitemShut {NoStop}%
\bibitem [{\citenamefont {Mahmood}\ \emph {et~al.}(2016)\citenamefont
  {Mahmood}, \citenamefont {Chan}, \citenamefont {Alpichshev}, \citenamefont
  {Gardner}, \citenamefont {Lee}, \citenamefont {Lee},\ and\ \citenamefont
  {Gedik}}]{Mahmood2016}%
  \BibitemOpen
  \bibfield  {author} {\bibinfo {author} {\bibfnamefont {F.}~\bibnamefont
  {Mahmood}}, \bibinfo {author} {\bibfnamefont {C.-K.}\ \bibnamefont {Chan}},
  \bibinfo {author} {\bibfnamefont {Z.}~\bibnamefont {Alpichshev}}, \bibinfo
  {author} {\bibfnamefont {D.}~\bibnamefont {Gardner}}, \bibinfo {author}
  {\bibfnamefont {Y.}~\bibnamefont {Lee}}, \bibinfo {author} {\bibfnamefont
  {P.~A.}\ \bibnamefont {Lee}}, \ and\ \bibinfo {author} {\bibfnamefont
  {N.}~\bibnamefont {Gedik}},\ }\href {\doibase 10.1038/nphys3609} {\bibfield
  {journal} {\bibinfo  {journal} {Nature Physics}\ }\textbf {\bibinfo {volume}
  {12}},\ \bibinfo {pages} {306} (\bibinfo {year} {2016})}\BibitemShut
  {NoStop}%
\bibitem [{\citenamefont {Titum}\ \emph {et~al.}(2016)\citenamefont {Titum},
  \citenamefont {Berg}, \citenamefont {Rudner}, \citenamefont {Refael},\ and\
  \citenamefont {Lindner}}]{Titum2016}%
  \BibitemOpen
  \bibfield  {author} {\bibinfo {author} {\bibfnamefont {P.}~\bibnamefont
  {Titum}}, \bibinfo {author} {\bibfnamefont {E.}~\bibnamefont {Berg}},
  \bibinfo {author} {\bibfnamefont {M.~S.}\ \bibnamefont {Rudner}}, \bibinfo
  {author} {\bibfnamefont {G.}~\bibnamefont {Refael}}, \ and\ \bibinfo {author}
  {\bibfnamefont {N.~H.}\ \bibnamefont {Lindner}},\ }\href {\doibase
  10.1103/PhysRevX.6.021013} {\bibfield  {journal} {\bibinfo  {journal}
  {Physical Review X}\ }\textbf {\bibinfo {volume} {6}},\ \bibinfo {pages}
  {021013} (\bibinfo {year} {2016})}\BibitemShut {NoStop}%
\bibitem [{\citenamefont {Nathan}\ \emph {et~al.}(2016)\citenamefont {Nathan},
  \citenamefont {Rudner}, \citenamefont {Lindner}, \citenamefont {Berg},\ and\
  \citenamefont {Refael}}]{Nathan2016}%
  \BibitemOpen
  \bibfield  {author} {\bibinfo {author} {\bibfnamefont {F.}~\bibnamefont
  {Nathan}}, \bibinfo {author} {\bibfnamefont {M.~S.}\ \bibnamefont {Rudner}},
  \bibinfo {author} {\bibfnamefont {N.~H.}\ \bibnamefont {Lindner}}, \bibinfo
  {author} {\bibfnamefont {E.}~\bibnamefont {Berg}}, \ and\ \bibinfo {author}
  {\bibfnamefont {G.}~\bibnamefont {Refael}},\ }\href
  {http://arxiv.org/abs/1610.03590} {\  (\bibinfo {year} {2016})},\ \Eprint
  {http://arxiv.org/abs/1610.03590} {arXiv:1610.03590} \BibitemShut {NoStop}%
\bibitem [{\citenamefont {Maczewsky}\ \emph {et~al.}(2017)\citenamefont
  {Maczewsky}, \citenamefont {Zeuner}, \citenamefont {Nolte},\ and\
  \citenamefont {Szameit}}]{Maczewsky2017}%
  \BibitemOpen
  \bibfield  {author} {\bibinfo {author} {\bibfnamefont {L.~J.}\ \bibnamefont
  {Maczewsky}}, \bibinfo {author} {\bibfnamefont {J.~M.}\ \bibnamefont
  {Zeuner}}, \bibinfo {author} {\bibfnamefont {S.}~\bibnamefont {Nolte}}, \
  and\ \bibinfo {author} {\bibfnamefont {A.}~\bibnamefont {Szameit}},\ }\href
  {\doibase 10.1038/ncomms13756} {\bibfield  {journal} {\bibinfo  {journal}
  {Nature Communications}\ }\textbf {\bibinfo {volume} {8}},\ \bibinfo {pages}
  {13756} (\bibinfo {year} {2017})}\BibitemShut {NoStop}%
\bibitem [{\citenamefont {Oka}\ and\ \citenamefont {Kitamura}(2019)}]{Oka2019}%
  \BibitemOpen
  \bibfield  {author} {\bibinfo {author} {\bibfnamefont {T.}~\bibnamefont
  {Oka}}\ and\ \bibinfo {author} {\bibfnamefont {S.}~\bibnamefont {Kitamura}},\
  }\href {\doibase 10.1146/annurev-conmatphys-031218-013423} {\bibfield
  {journal} {\bibinfo  {journal} {Annual Review of Condensed Matter Physics}\
  }\textbf {\bibinfo {volume} {10}},\ \bibinfo {pages} {annurev} (\bibinfo
  {year} {2019})}\BibitemShut {NoStop}%
\bibitem [{\citenamefont {Kimel}\ \emph {et~al.}(2005)\citenamefont {Kimel},
  \citenamefont {Kirilyuk}, \citenamefont {Usachev}, \citenamefont {Pisarev},
  \citenamefont {Balbashov},\ and\ \citenamefont {Rasing}}]{Kimel2005}%
  \BibitemOpen
  \bibfield  {author} {\bibinfo {author} {\bibfnamefont {A.~V.}\ \bibnamefont
  {Kimel}}, \bibinfo {author} {\bibfnamefont {A.}~\bibnamefont {Kirilyuk}},
  \bibinfo {author} {\bibfnamefont {P.~A.}\ \bibnamefont {Usachev}}, \bibinfo
  {author} {\bibfnamefont {R.~V.}\ \bibnamefont {Pisarev}}, \bibinfo {author}
  {\bibfnamefont {A.~M.}\ \bibnamefont {Balbashov}}, \ and\ \bibinfo {author}
  {\bibfnamefont {T.}~\bibnamefont {Rasing}},\ }\href {\doibase
  10.1038/nature03564} {\bibfield  {journal} {\bibinfo  {journal} {Nature}\
  }\textbf {\bibinfo {volume} {435}},\ \bibinfo {pages} {655} (\bibinfo {year}
  {2005})}\BibitemShut {NoStop}%
\bibitem [{\citenamefont {Stanciu}\ \emph {et~al.}(2007)\citenamefont
  {Stanciu}, \citenamefont {Hansteen}, \citenamefont {Kimel}, \citenamefont
  {Kirilyuk}, \citenamefont {Tsukamoto}, \citenamefont {Itoh},\ and\
  \citenamefont {Rasing}}]{Stanciu2007}%
  \BibitemOpen
  \bibfield  {author} {\bibinfo {author} {\bibfnamefont {C.~D.}\ \bibnamefont
  {Stanciu}}, \bibinfo {author} {\bibfnamefont {F.}~\bibnamefont {Hansteen}},
  \bibinfo {author} {\bibfnamefont {A.~V.}\ \bibnamefont {Kimel}}, \bibinfo
  {author} {\bibfnamefont {A.}~\bibnamefont {Kirilyuk}}, \bibinfo {author}
  {\bibfnamefont {A.}~\bibnamefont {Tsukamoto}}, \bibinfo {author}
  {\bibfnamefont {A.}~\bibnamefont {Itoh}}, \ and\ \bibinfo {author}
  {\bibfnamefont {T.}~\bibnamefont {Rasing}},\ }\href {\doibase
  10.1103/PhysRevLett.99.047601} {\bibfield  {journal} {\bibinfo  {journal}
  {Physical Review Letters}\ }\textbf {\bibinfo {volume} {99}},\ \bibinfo
  {pages} {047601} (\bibinfo {year} {2007})}\BibitemShut {NoStop}%
\bibitem [{\citenamefont {Yao}\ \emph {et~al.}(2007)\citenamefont {Yao},
  \citenamefont {MacDonald},\ and\ \citenamefont {Niu}}]{Yao2007}%
  \BibitemOpen
  \bibfield  {author} {\bibinfo {author} {\bibfnamefont {W.}~\bibnamefont
  {Yao}}, \bibinfo {author} {\bibfnamefont {A.~H.}\ \bibnamefont {MacDonald}},
  \ and\ \bibinfo {author} {\bibfnamefont {Q.}~\bibnamefont {Niu}},\ }\href
  {\doibase 10.1103/PhysRevLett.99.047401} {\bibfield  {journal} {\bibinfo
  {journal} {Physical Review Letters}\ }\textbf {\bibinfo {volume} {99}},\
  \bibinfo {pages} {047401} (\bibinfo {year} {2007})}\BibitemShut {NoStop}%
\bibitem [{\citenamefont {Oka}\ and\ \citenamefont {Aoki}(2009)}]{Oka2009}%
  \BibitemOpen
  \bibfield  {author} {\bibinfo {author} {\bibfnamefont {T.}~\bibnamefont
  {Oka}}\ and\ \bibinfo {author} {\bibfnamefont {H.}~\bibnamefont {Aoki}},\
  }\href {\doibase 10.1103/PhysRevB.79.081406} {\bibfield  {journal} {\bibinfo
  {journal} {Physical Review B}\ }\textbf {\bibinfo {volume} {79}},\ \bibinfo
  {pages} {081406} (\bibinfo {year} {2009})}\BibitemShut {NoStop}%
\bibitem [{\citenamefont {Kirilyuk}\ \emph {et~al.}(2010)\citenamefont
  {Kirilyuk}, \citenamefont {Kimel},\ and\ \citenamefont
  {Rasing}}]{Kirilyuk2010}%
  \BibitemOpen
  \bibfield  {author} {\bibinfo {author} {\bibfnamefont {A.}~\bibnamefont
  {Kirilyuk}}, \bibinfo {author} {\bibfnamefont {A.~V.}\ \bibnamefont {Kimel}},
  \ and\ \bibinfo {author} {\bibfnamefont {T.}~\bibnamefont {Rasing}},\ }\href
  {\doibase 10.1103/RevModPhys.82.2731} {\bibfield  {journal} {\bibinfo
  {journal} {Reviews of Modern Physics}\ }\textbf {\bibinfo {volume} {82}},\
  \bibinfo {pages} {2731} (\bibinfo {year} {2010})}\BibitemShut {NoStop}%
\bibitem [{\citenamefont {Fausti}\ \emph {et~al.}(2011)\citenamefont {Fausti},
  \citenamefont {Tobey}, \citenamefont {Dean}, \citenamefont {Kaiser},
  \citenamefont {Dienst}, \citenamefont {Hoffmann}, \citenamefont {Pyon},
  \citenamefont {Takayama}, \citenamefont {Takagi},\ and\ \citenamefont
  {Cavalleri}}]{Fausti2011}%
  \BibitemOpen
  \bibfield  {author} {\bibinfo {author} {\bibfnamefont {D.}~\bibnamefont
  {Fausti}}, \bibinfo {author} {\bibfnamefont {R.~I.}\ \bibnamefont {Tobey}},
  \bibinfo {author} {\bibfnamefont {N.}~\bibnamefont {Dean}}, \bibinfo {author}
  {\bibfnamefont {S.}~\bibnamefont {Kaiser}}, \bibinfo {author} {\bibfnamefont
  {A.}~\bibnamefont {Dienst}}, \bibinfo {author} {\bibfnamefont {M.~C.}\
  \bibnamefont {Hoffmann}}, \bibinfo {author} {\bibfnamefont {S.}~\bibnamefont
  {Pyon}}, \bibinfo {author} {\bibfnamefont {T.}~\bibnamefont {Takayama}},
  \bibinfo {author} {\bibfnamefont {H.}~\bibnamefont {Takagi}}, \ and\ \bibinfo
  {author} {\bibfnamefont {A.}~\bibnamefont {Cavalleri}},\ }\href {\doibase
  10.1126/science.1197294} {\bibfield  {journal} {\bibinfo  {journal} {Science
  (New York, N.Y.)}\ }\textbf {\bibinfo {volume} {331}},\ \bibinfo {pages}
  {189} (\bibinfo {year} {2011})}\BibitemShut {NoStop}%
\bibitem [{\citenamefont {Lindner}\ \emph {et~al.}(2011)\citenamefont
  {Lindner}, \citenamefont {Refael},\ and\ \citenamefont
  {Galitski}}]{Lindner2011}%
  \BibitemOpen
  \bibfield  {author} {\bibinfo {author} {\bibfnamefont {N.~H.}\ \bibnamefont
  {Lindner}}, \bibinfo {author} {\bibfnamefont {G.}~\bibnamefont {Refael}}, \
  and\ \bibinfo {author} {\bibfnamefont {V.}~\bibnamefont {Galitski}},\ }\href
  {\doibase 10.1038/nphys1926} {\bibfield  {journal} {\bibinfo  {journal}
  {Nature Physics}\ }\textbf {\bibinfo {volume} {7}},\ \bibinfo {pages} {490}
  (\bibinfo {year} {2011})}\BibitemShut {NoStop}%
\bibitem [{\citenamefont {Lindner}\ \emph {et~al.}(2013)\citenamefont
  {Lindner}, \citenamefont {Bergman}, \citenamefont {Refael},\ and\
  \citenamefont {Galitski}}]{Lindner2013}%
  \BibitemOpen
  \bibfield  {author} {\bibinfo {author} {\bibfnamefont {N.~H.}\ \bibnamefont
  {Lindner}}, \bibinfo {author} {\bibfnamefont {D.~L.}\ \bibnamefont
  {Bergman}}, \bibinfo {author} {\bibfnamefont {G.}~\bibnamefont {Refael}}, \
  and\ \bibinfo {author} {\bibfnamefont {V.}~\bibnamefont {Galitski}},\ }\href
  {\doibase 10.1103/PhysRevB.87.235131} {\bibfield  {journal} {\bibinfo
  {journal} {Physical Review B}\ }\textbf {\bibinfo {volume} {87}},\ \bibinfo
  {pages} {235131} (\bibinfo {year} {2013})}\BibitemShut {NoStop}%
\bibitem [{\citenamefont {Katan}\ and\ \citenamefont
  {Podolsky}(2013)}]{Katan2013}%
  \BibitemOpen
  \bibfield  {author} {\bibinfo {author} {\bibfnamefont {Y.~T.}\ \bibnamefont
  {Katan}}\ and\ \bibinfo {author} {\bibfnamefont {D.}~\bibnamefont
  {Podolsky}},\ }\href {\doibase 10.1103/PhysRevLett.110.016802} {\bibfield
  {journal} {\bibinfo  {journal} {Physical Review Letters}\ }\textbf {\bibinfo
  {volume} {110}},\ \bibinfo {pages} {016802} (\bibinfo {year}
  {2013})}\BibitemShut {NoStop}%
\bibitem [{\citenamefont {Cayssol}\ \emph {et~al.}(2013)\citenamefont
  {Cayssol}, \citenamefont {D{\'{o}}ra}, \citenamefont {Simon},\ and\
  \citenamefont {Moessner}}]{Cayssol2013}%
  \BibitemOpen
  \bibfield  {author} {\bibinfo {author} {\bibfnamefont {J.}~\bibnamefont
  {Cayssol}}, \bibinfo {author} {\bibfnamefont {B.}~\bibnamefont {D{\'{o}}ra}},
  \bibinfo {author} {\bibfnamefont {F.}~\bibnamefont {Simon}}, \ and\ \bibinfo
  {author} {\bibfnamefont {R.}~\bibnamefont {Moessner}},\ }\href {\doibase
  10.1002/pssr.201206451} {\bibfield  {journal} {\bibinfo  {journal} {physica
  status solidi (RRL) - Rapid Research Letters}\ }\textbf {\bibinfo {volume}
  {7}},\ \bibinfo {pages} {101} (\bibinfo {year} {2013})}\BibitemShut {NoStop}%
\bibitem [{\citenamefont {Hu}\ \emph {et~al.}(2014)\citenamefont {Hu},
  \citenamefont {Kaiser}, \citenamefont {Nicoletti}, \citenamefont {Hunt},
  \citenamefont {Gierz}, \citenamefont {Hoffmann}, \citenamefont {{Le Tacon}},
  \citenamefont {Loew}, \citenamefont {Keimer},\ and\ \citenamefont
  {Cavalleri}}]{Hu2014}%
  \BibitemOpen
  \bibfield  {author} {\bibinfo {author} {\bibfnamefont {W.}~\bibnamefont
  {Hu}}, \bibinfo {author} {\bibfnamefont {S.}~\bibnamefont {Kaiser}}, \bibinfo
  {author} {\bibfnamefont {D.}~\bibnamefont {Nicoletti}}, \bibinfo {author}
  {\bibfnamefont {C.~R.}\ \bibnamefont {Hunt}}, \bibinfo {author}
  {\bibfnamefont {I.}~\bibnamefont {Gierz}}, \bibinfo {author} {\bibfnamefont
  {M.~C.}\ \bibnamefont {Hoffmann}}, \bibinfo {author} {\bibfnamefont
  {M.}~\bibnamefont {{Le Tacon}}}, \bibinfo {author} {\bibfnamefont
  {T.}~\bibnamefont {Loew}}, \bibinfo {author} {\bibfnamefont {B.}~\bibnamefont
  {Keimer}}, \ and\ \bibinfo {author} {\bibfnamefont {A.}~\bibnamefont
  {Cavalleri}},\ }\href {\doibase 10.1038/nmat3963} {\bibfield  {journal}
  {\bibinfo  {journal} {Nature Materials}\ }\textbf {\bibinfo {volume} {13}},\
  \bibinfo {pages} {705} (\bibinfo {year} {2014})}\BibitemShut {NoStop}%
\bibitem [{\citenamefont {Mitrano}\ \emph {et~al.}(2016)\citenamefont
  {Mitrano}, \citenamefont {Cantaluppi}, \citenamefont {Nicoletti},
  \citenamefont {Kaiser}, \citenamefont {Perucchi}, \citenamefont {Lupi},
  \citenamefont {{Di Pietro}}, \citenamefont {Pontiroli}, \citenamefont
  {Ricc{\`{o}}}, \citenamefont {Clark}, \citenamefont {Jaksch},\ and\
  \citenamefont {Cavalleri}}]{Mitrano2016}%
  \BibitemOpen
  \bibfield  {author} {\bibinfo {author} {\bibfnamefont {M.}~\bibnamefont
  {Mitrano}}, \bibinfo {author} {\bibfnamefont {A.}~\bibnamefont {Cantaluppi}},
  \bibinfo {author} {\bibfnamefont {D.}~\bibnamefont {Nicoletti}}, \bibinfo
  {author} {\bibfnamefont {S.}~\bibnamefont {Kaiser}}, \bibinfo {author}
  {\bibfnamefont {A.}~\bibnamefont {Perucchi}}, \bibinfo {author}
  {\bibfnamefont {S.}~\bibnamefont {Lupi}}, \bibinfo {author} {\bibfnamefont
  {P.}~\bibnamefont {{Di Pietro}}}, \bibinfo {author} {\bibfnamefont
  {D.}~\bibnamefont {Pontiroli}}, \bibinfo {author} {\bibfnamefont
  {M.}~\bibnamefont {Ricc{\`{o}}}}, \bibinfo {author} {\bibfnamefont {S.~R.}\
  \bibnamefont {Clark}}, \bibinfo {author} {\bibfnamefont {D.}~\bibnamefont
  {Jaksch}}, \ and\ \bibinfo {author} {\bibfnamefont {A.}~\bibnamefont
  {Cavalleri}},\ }\href {\doibase 10.1038/nature16522} {\bibfield  {journal}
  {\bibinfo  {journal} {Nature}\ }\textbf {\bibinfo {volume} {530}},\ \bibinfo
  {pages} {461} (\bibinfo {year} {2016})}\BibitemShut {NoStop}%
\bibitem [{\citenamefont {Cavalleri}(2018)}]{Cavalleri2018}%
  \BibitemOpen
  \bibfield  {author} {\bibinfo {author} {\bibfnamefont {A.}~\bibnamefont
  {Cavalleri}},\ }\href {\doibase 10.1080/00107514.2017.1406623} {\bibfield
  {journal} {\bibinfo  {journal} {Contemporary Physics}\ }\textbf {\bibinfo
  {volume} {59}},\ \bibinfo {pages} {31} (\bibinfo {year} {2018})}\BibitemShut
  {NoStop}%
\bibitem [{\citenamefont {Gu}\ \emph {et~al.}(2011)\citenamefont {Gu},
  \citenamefont {Fertig}, \citenamefont {Arovas},\ and\ \citenamefont
  {Auerbach}}]{Gu2011}%
  \BibitemOpen
  \bibfield  {author} {\bibinfo {author} {\bibfnamefont {Z.}~\bibnamefont
  {Gu}}, \bibinfo {author} {\bibfnamefont {H.~A.}\ \bibnamefont {Fertig}},
  \bibinfo {author} {\bibfnamefont {D.~P.}\ \bibnamefont {Arovas}}, \ and\
  \bibinfo {author} {\bibfnamefont {A.}~\bibnamefont {Auerbach}},\ }\href
  {\doibase 10.1103/PhysRevLett.107.216601} {\bibfield  {journal} {\bibinfo
  {journal} {Physical Review Letters}\ }\textbf {\bibinfo {volume} {107}},\
  \bibinfo {pages} {216601} (\bibinfo {year} {2011})}\BibitemShut {NoStop}%
\bibitem [{\citenamefont {Kitagawa}\ \emph {et~al.}(2011)\citenamefont
  {Kitagawa}, \citenamefont {Oka}, \citenamefont {Brataas}, \citenamefont
  {Fu},\ and\ \citenamefont {Demler}}]{Kitagawa2011}%
  \BibitemOpen
  \bibfield  {author} {\bibinfo {author} {\bibfnamefont {T.}~\bibnamefont
  {Kitagawa}}, \bibinfo {author} {\bibfnamefont {T.}~\bibnamefont {Oka}},
  \bibinfo {author} {\bibfnamefont {A.}~\bibnamefont {Brataas}}, \bibinfo
  {author} {\bibfnamefont {L.}~\bibnamefont {Fu}}, \ and\ \bibinfo {author}
  {\bibfnamefont {E.}~\bibnamefont {Demler}},\ }\href {\doibase
  10.1103/PhysRevB.84.235108} {\bibfield  {journal} {\bibinfo  {journal}
  {Physical Review B}\ }\textbf {\bibinfo {volume} {84}},\ \bibinfo {pages}
  {235108} (\bibinfo {year} {2011})}\BibitemShut {NoStop}%
\bibitem [{\citenamefont {D'Alessio}\ and\ \citenamefont
  {Rigol}(2014)}]{DAlessio2014}%
  \BibitemOpen
  \bibfield  {author} {\bibinfo {author} {\bibfnamefont {L.}~\bibnamefont
  {D'Alessio}}\ and\ \bibinfo {author} {\bibfnamefont {M.}~\bibnamefont
  {Rigol}},\ }\href {\doibase 10.1103/PhysRevX.4.041048} {\bibfield  {journal}
  {\bibinfo  {journal} {Physical Review X}\ }\textbf {\bibinfo {volume} {4}},\
  \bibinfo {pages} {041048} (\bibinfo {year} {2014})}\BibitemShut {NoStop}%
\bibitem [{\citenamefont {Iadecola}\ \emph {et~al.}(2013)\citenamefont
  {Iadecola}, \citenamefont {Campbell}, \citenamefont {Chamon}, \citenamefont
  {Hou}, \citenamefont {Jackiw}, \citenamefont {Pi},\ and\ \citenamefont
  {Kusminskiy}}]{Iadecola2013}%
  \BibitemOpen
  \bibfield  {author} {\bibinfo {author} {\bibfnamefont {T.}~\bibnamefont
  {Iadecola}}, \bibinfo {author} {\bibfnamefont {D.}~\bibnamefont {Campbell}},
  \bibinfo {author} {\bibfnamefont {C.}~\bibnamefont {Chamon}}, \bibinfo
  {author} {\bibfnamefont {C.-Y.}\ \bibnamefont {Hou}}, \bibinfo {author}
  {\bibfnamefont {R.}~\bibnamefont {Jackiw}}, \bibinfo {author} {\bibfnamefont
  {S.-Y.}\ \bibnamefont {Pi}}, \ and\ \bibinfo {author} {\bibfnamefont {S.~V.}\
  \bibnamefont {Kusminskiy}},\ }\href {\doibase 10.1103/PhysRevLett.110.176603}
  {\bibfield  {journal} {\bibinfo  {journal} {Physical Review Letters}\
  }\textbf {\bibinfo {volume} {110}},\ \bibinfo {pages} {176603} (\bibinfo
  {year} {2013})}\BibitemShut {NoStop}%
\bibitem [{\citenamefont {Dehghani}\ \emph {et~al.}(2014)\citenamefont
  {Dehghani}, \citenamefont {Oka},\ and\ \citenamefont {Mitra}}]{Dehghani2014}%
  \BibitemOpen
  \bibfield  {author} {\bibinfo {author} {\bibfnamefont {H.}~\bibnamefont
  {Dehghani}}, \bibinfo {author} {\bibfnamefont {T.}~\bibnamefont {Oka}}, \
  and\ \bibinfo {author} {\bibfnamefont {A.}~\bibnamefont {Mitra}},\ }\href
  {\doibase 10.1103/PhysRevB.90.195429} {\bibfield  {journal} {\bibinfo
  {journal} {Physical Review B}\ }\textbf {\bibinfo {volume} {90}},\ \bibinfo
  {pages} {195429} (\bibinfo {year} {2014})}\BibitemShut {NoStop}%
\bibitem [{\citenamefont {Dehghani}\ \emph {et~al.}(2015)\citenamefont
  {Dehghani}, \citenamefont {Oka},\ and\ \citenamefont {Mitra}}]{Dehghani2015}%
  \BibitemOpen
  \bibfield  {author} {\bibinfo {author} {\bibfnamefont {H.}~\bibnamefont
  {Dehghani}}, \bibinfo {author} {\bibfnamefont {T.}~\bibnamefont {Oka}}, \
  and\ \bibinfo {author} {\bibfnamefont {A.}~\bibnamefont {Mitra}},\ }\href
  {\doibase 10.1103/PhysRevB.91.155422} {\bibfield  {journal} {\bibinfo
  {journal} {Physical Review B}\ }\textbf {\bibinfo {volume} {91}},\ \bibinfo
  {pages} {155422} (\bibinfo {year} {2015})}\BibitemShut {NoStop}%
\bibitem [{\citenamefont {Iadecola}\ and\ \citenamefont
  {Chamon}(2015)}]{Iadecola2015}%
  \BibitemOpen
  \bibfield  {author} {\bibinfo {author} {\bibfnamefont {T.}~\bibnamefont
  {Iadecola}}\ and\ \bibinfo {author} {\bibfnamefont {C.}~\bibnamefont
  {Chamon}},\ }\href {\doibase 10.1103/PhysRevB.91.184301} {\bibfield
  {journal} {\bibinfo  {journal} {Physical Review B}\ }\textbf {\bibinfo
  {volume} {91}},\ \bibinfo {pages} {184301} (\bibinfo {year}
  {2015})}\BibitemShut {NoStop}%
\bibitem [{\citenamefont {Iadecola}\ \emph {et~al.}(2015)\citenamefont
  {Iadecola}, \citenamefont {Neupert},\ and\ \citenamefont
  {Chamon}}]{Iadecola2015a}%
  \BibitemOpen
  \bibfield  {author} {\bibinfo {author} {\bibfnamefont {T.}~\bibnamefont
  {Iadecola}}, \bibinfo {author} {\bibfnamefont {T.}~\bibnamefont {Neupert}}, \
  and\ \bibinfo {author} {\bibfnamefont {C.}~\bibnamefont {Chamon}},\ }\href
  {\doibase 10.1103/PhysRevB.91.235133} {\bibfield  {journal} {\bibinfo
  {journal} {Physical Review B}\ }\textbf {\bibinfo {volume} {91}},\ \bibinfo
  {pages} {235133} (\bibinfo {year} {2015})}\BibitemShut {NoStop}%
\bibitem [{\citenamefont {Shirai}\ \emph {et~al.}(2015)\citenamefont {Shirai},
  \citenamefont {Mori},\ and\ \citenamefont {Miyashita}}]{Shirai2015}%
  \BibitemOpen
  \bibfield  {author} {\bibinfo {author} {\bibfnamefont {T.}~\bibnamefont
  {Shirai}}, \bibinfo {author} {\bibfnamefont {T.}~\bibnamefont {Mori}}, \ and\
  \bibinfo {author} {\bibfnamefont {S.}~\bibnamefont {Miyashita}},\ }\href
  {\doibase 10.1103/PhysRevE.91.030101} {\bibfield  {journal} {\bibinfo
  {journal} {Physical Review E}\ }\textbf {\bibinfo {volume} {91}},\ \bibinfo
  {pages} {030101} (\bibinfo {year} {2015})}\BibitemShut {NoStop}%
\bibitem [{\citenamefont {Liu}(2015)}]{Liu2015}%
  \BibitemOpen
  \bibfield  {author} {\bibinfo {author} {\bibfnamefont {D.~E.}\ \bibnamefont
  {Liu}},\ }\href {\doibase 10.1103/PhysRevB.91.144301} {\bibfield  {journal}
  {\bibinfo  {journal} {Physical Review B}\ }\textbf {\bibinfo {volume} {91}},\
  \bibinfo {pages} {144301} (\bibinfo {year} {2015})}\BibitemShut {NoStop}%
\bibitem [{\citenamefont {Dehghani}\ and\ \citenamefont
  {Mitra}(2016)}]{Dehghani2016}%
  \BibitemOpen
  \bibfield  {author} {\bibinfo {author} {\bibfnamefont {H.}~\bibnamefont
  {Dehghani}}\ and\ \bibinfo {author} {\bibfnamefont {A.}~\bibnamefont
  {Mitra}},\ }\href {\doibase 10.1103/PhysRevB.93.205437} {\bibfield  {journal}
  {\bibinfo  {journal} {Physical Review B}\ }\textbf {\bibinfo {volume} {93}},\
  \bibinfo {pages} {205437} (\bibinfo {year} {2016})}\BibitemShut {NoStop}%
\bibitem [{\citenamefont {Else}\ \emph {et~al.}(2016)\citenamefont {Else},
  \citenamefont {Bauer},\ and\ \citenamefont {Nayak}}]{Else2016}%
  \BibitemOpen
  \bibfield  {author} {\bibinfo {author} {\bibfnamefont {D.~V.}\ \bibnamefont
  {Else}}, \bibinfo {author} {\bibfnamefont {B.}~\bibnamefont {Bauer}}, \ and\
  \bibinfo {author} {\bibfnamefont {C.}~\bibnamefont {Nayak}},\ }\href
  {\doibase 10.1103/PhysRevLett.117.090402} {\bibfield  {journal} {\bibinfo
  {journal} {Physical Review Letters}\ }\textbf {\bibinfo {volume} {117}},\
  \bibinfo {pages} {090402} (\bibinfo {year} {2016})}\BibitemShut {NoStop}%
\bibitem [{\citenamefont {Khemani}\ \emph {et~al.}(2016)\citenamefont
  {Khemani}, \citenamefont {Lazarides}, \citenamefont {Moessner},\ and\
  \citenamefont {Sondhi}}]{Khemani2016}%
  \BibitemOpen
  \bibfield  {author} {\bibinfo {author} {\bibfnamefont {V.}~\bibnamefont
  {Khemani}}, \bibinfo {author} {\bibfnamefont {A.}~\bibnamefont {Lazarides}},
  \bibinfo {author} {\bibfnamefont {R.}~\bibnamefont {Moessner}}, \ and\
  \bibinfo {author} {\bibfnamefont {S.}~\bibnamefont {Sondhi}},\ }\href
  {\doibase 10.1103/PhysRevLett.116.250401} {\bibfield  {journal} {\bibinfo
  {journal} {Physical Review Letters}\ }\textbf {\bibinfo {volume} {116}},\
  \bibinfo {pages} {250401} (\bibinfo {year} {2016})}\BibitemShut {NoStop}%
\bibitem [{\citenamefont {Yao}\ \emph {et~al.}(2017)\citenamefont {Yao},
  \citenamefont {Potter}, \citenamefont {Potirniche},\ and\ \citenamefont
  {Vishwanath}}]{Yao2017}%
  \BibitemOpen
  \bibfield  {author} {\bibinfo {author} {\bibfnamefont {N.}~\bibnamefont
  {Yao}}, \bibinfo {author} {\bibfnamefont {A.}~\bibnamefont {Potter}},
  \bibinfo {author} {\bibfnamefont {I.-D.}\ \bibnamefont {Potirniche}}, \ and\
  \bibinfo {author} {\bibfnamefont {A.}~\bibnamefont {Vishwanath}},\ }\href
  {\doibase 10.1103/PhysRevLett.118.030401} {\bibfield  {journal} {\bibinfo
  {journal} {Physical Review Letters}\ }\textbf {\bibinfo {volume} {118}},\
  \bibinfo {pages} {030401} (\bibinfo {year} {2017})}\BibitemShut {NoStop}%
\bibitem [{\citenamefont {Choi}\ \emph {et~al.}(2017)\citenamefont {Choi},
  \citenamefont {Choi}, \citenamefont {Landig}, \citenamefont {Kucsko},
  \citenamefont {Zhou}, \citenamefont {Isoya}, \citenamefont {Jelezko},
  \citenamefont {Onoda}, \citenamefont {Sumiya}, \citenamefont {Khemani},
  \citenamefont {von Keyserlingk}, \citenamefont {Yao}, \citenamefont
  {Demler},\ and\ \citenamefont {Lukin}}]{Choi2017}%
  \BibitemOpen
  \bibfield  {author} {\bibinfo {author} {\bibfnamefont {S.}~\bibnamefont
  {Choi}}, \bibinfo {author} {\bibfnamefont {J.}~\bibnamefont {Choi}}, \bibinfo
  {author} {\bibfnamefont {R.}~\bibnamefont {Landig}}, \bibinfo {author}
  {\bibfnamefont {G.}~\bibnamefont {Kucsko}}, \bibinfo {author} {\bibfnamefont
  {H.}~\bibnamefont {Zhou}}, \bibinfo {author} {\bibfnamefont {J.}~\bibnamefont
  {Isoya}}, \bibinfo {author} {\bibfnamefont {F.}~\bibnamefont {Jelezko}},
  \bibinfo {author} {\bibfnamefont {S.}~\bibnamefont {Onoda}}, \bibinfo
  {author} {\bibfnamefont {H.}~\bibnamefont {Sumiya}}, \bibinfo {author}
  {\bibfnamefont {V.}~\bibnamefont {Khemani}}, \bibinfo {author} {\bibfnamefont
  {C.}~\bibnamefont {von Keyserlingk}}, \bibinfo {author} {\bibfnamefont
  {N.~Y.}\ \bibnamefont {Yao}}, \bibinfo {author} {\bibfnamefont
  {E.}~\bibnamefont {Demler}}, \ and\ \bibinfo {author} {\bibfnamefont {M.~D.}\
  \bibnamefont {Lukin}},\ }\href {http://dx.doi.org/10.1038/nature21426
  http://10.0.4.14/nature21426} {\bibfield  {journal} {\bibinfo  {journal}
  {Nature}\ }\textbf {\bibinfo {volume} {543}},\ \bibinfo {pages} {221}
  (\bibinfo {year} {2017})}\BibitemShut {NoStop}%
\bibitem [{\citenamefont {Zhang}\ \emph {et~al.}(2017)\citenamefont {Zhang},
  \citenamefont {Hess}, \citenamefont {Kyprianidis}, \citenamefont {Becker},
  \citenamefont {Lee}, \citenamefont {Smith}, \citenamefont {Pagano},
  \citenamefont {Potirniche}, \citenamefont {Potter}, \citenamefont
  {Vishwanath}, \citenamefont {Yao},\ and\ \citenamefont {Monroe}}]{Zhang2017}%
  \BibitemOpen
  \bibfield  {author} {\bibinfo {author} {\bibfnamefont {J.}~\bibnamefont
  {Zhang}}, \bibinfo {author} {\bibfnamefont {P.~W.}\ \bibnamefont {Hess}},
  \bibinfo {author} {\bibfnamefont {A.}~\bibnamefont {Kyprianidis}}, \bibinfo
  {author} {\bibfnamefont {P.}~\bibnamefont {Becker}}, \bibinfo {author}
  {\bibfnamefont {A.}~\bibnamefont {Lee}}, \bibinfo {author} {\bibfnamefont
  {J.}~\bibnamefont {Smith}}, \bibinfo {author} {\bibfnamefont
  {G.}~\bibnamefont {Pagano}}, \bibinfo {author} {\bibfnamefont {I.-D.}\
  \bibnamefont {Potirniche}}, \bibinfo {author} {\bibfnamefont {A.~C.}\
  \bibnamefont {Potter}}, \bibinfo {author} {\bibfnamefont {A.}~\bibnamefont
  {Vishwanath}}, \bibinfo {author} {\bibfnamefont {N.~Y.}\ \bibnamefont {Yao}},
  \ and\ \bibinfo {author} {\bibfnamefont {C.}~\bibnamefont {Monroe}},\ }\href
  {http://dx.doi.org/10.1038/nature21413 http://10.0.4.14/nature21413}
  {\bibfield  {journal} {\bibinfo  {journal} {Nature}\ }\textbf {\bibinfo
  {volume} {543}},\ \bibinfo {pages} {217} (\bibinfo {year}
  {2017})}\BibitemShut {NoStop}%
\bibitem [{\citenamefont {Esin}\ \emph {et~al.}(2018)\citenamefont {Esin},
  \citenamefont {Rudner}, \citenamefont {Refael},\ and\ \citenamefont
  {Lindner}}]{Esin2018}%
  \BibitemOpen
  \bibfield  {author} {\bibinfo {author} {\bibfnamefont {I.}~\bibnamefont
  {Esin}}, \bibinfo {author} {\bibfnamefont {M.~S.}\ \bibnamefont {Rudner}},
  \bibinfo {author} {\bibfnamefont {G.}~\bibnamefont {Refael}}, \ and\ \bibinfo
  {author} {\bibfnamefont {N.~H.}\ \bibnamefont {Lindner}},\ }\href {\doibase
  10.1103/PhysRevB.97.245401} {\bibfield  {journal} {\bibinfo  {journal}
  {Physical Review B}\ }\textbf {\bibinfo {volume} {97}},\ \bibinfo {pages}
  {245401} (\bibinfo {year} {2018})}\BibitemShut {NoStop}%
\bibitem [{\citenamefont {McIver}\ \emph {et~al.}(2018)\citenamefont {McIver},
  \citenamefont {Schulte}, \citenamefont {Stein}, \citenamefont {Matsuyama},
  \citenamefont {Jotzu}, \citenamefont {Meier},\ and\ \citenamefont
  {Cavalleri}}]{McIver2018}%
  \BibitemOpen
  \bibfield  {author} {\bibinfo {author} {\bibfnamefont {J.~W.}\ \bibnamefont
  {McIver}}, \bibinfo {author} {\bibfnamefont {B.}~\bibnamefont {Schulte}},
  \bibinfo {author} {\bibfnamefont {F.~U.}\ \bibnamefont {Stein}}, \bibinfo
  {author} {\bibfnamefont {T.}~\bibnamefont {Matsuyama}}, \bibinfo {author}
  {\bibfnamefont {G.}~\bibnamefont {Jotzu}}, \bibinfo {author} {\bibfnamefont
  {G.}~\bibnamefont {Meier}}, \ and\ \bibinfo {author} {\bibfnamefont
  {A.}~\bibnamefont {Cavalleri}},\ }\href {http://arxiv.org/abs/1811.03522} {\
  (\bibinfo {year} {2018})},\ \Eprint {http://arxiv.org/abs/1811.03522}
  {arXiv:1811.03522} \BibitemShut {NoStop}%
\bibitem [{\citenamefont {Seetharam}\ \emph {et~al.}(2015)\citenamefont
  {Seetharam}, \citenamefont {Bardyn}, \citenamefont {Lindner}, \citenamefont
  {Rudner},\ and\ \citenamefont {Refael}}]{Seetharam2015}%
  \BibitemOpen
  \bibfield  {author} {\bibinfo {author} {\bibfnamefont {K.~I.}\ \bibnamefont
  {Seetharam}}, \bibinfo {author} {\bibfnamefont {C.-E.}\ \bibnamefont
  {Bardyn}}, \bibinfo {author} {\bibfnamefont {N.~H.}\ \bibnamefont {Lindner}},
  \bibinfo {author} {\bibfnamefont {M.~S.}\ \bibnamefont {Rudner}}, \ and\
  \bibinfo {author} {\bibfnamefont {G.}~\bibnamefont {Refael}},\ }\href
  {\doibase 10.1103/PhysRevX.5.041050} {\bibfield  {journal} {\bibinfo
  {journal} {Physical Review X}\ }\textbf {\bibinfo {volume} {5}},\ \bibinfo
  {pages} {041050} (\bibinfo {year} {2015})}\BibitemShut {NoStop}%
\bibitem [{\citenamefont {Seetharam}\ \emph {et~al.}(2019)\citenamefont
  {Seetharam}, \citenamefont {Bardyn}, \citenamefont {Lindner}, \citenamefont
  {Rudner},\ and\ \citenamefont {Refael}}]{Seetharam2019}%
  \BibitemOpen
  \bibfield  {author} {\bibinfo {author} {\bibfnamefont {K.~I.}\ \bibnamefont
  {Seetharam}}, \bibinfo {author} {\bibfnamefont {C.-E.}\ \bibnamefont
  {Bardyn}}, \bibinfo {author} {\bibfnamefont {N.~H.}\ \bibnamefont {Lindner}},
  \bibinfo {author} {\bibfnamefont {M.~S.}\ \bibnamefont {Rudner}}, \ and\
  \bibinfo {author} {\bibfnamefont {G.}~\bibnamefont {Refael}},\ }\href
  {\doibase 10.1103/PhysRevB.99.014307} {\bibfield  {journal} {\bibinfo
  {journal} {Physical Review B}\ }\textbf {\bibinfo {volume} {99}},\ \bibinfo
  {pages} {014307} (\bibinfo {year} {2019})}\BibitemShut {NoStop}%
\bibitem [{\citenamefont {Dykman}\ \emph {et~al.}(2011)\citenamefont {Dykman},
  \citenamefont {Marthaler},\ and\ \citenamefont {Peano}}]{Dykman2011}%
  \BibitemOpen
  \bibfield  {author} {\bibinfo {author} {\bibfnamefont {M.~I.}\ \bibnamefont
  {Dykman}}, \bibinfo {author} {\bibfnamefont {M.}~\bibnamefont {Marthaler}}, \
  and\ \bibinfo {author} {\bibfnamefont {V.}~\bibnamefont {Peano}},\ }\href
  {\doibase 10.1103/PhysRevA.83.052115} {\bibfield  {journal} {\bibinfo
  {journal} {Physical Review A}\ }\textbf {\bibinfo {volume} {83}},\ \bibinfo
  {pages} {052115} (\bibinfo {year} {2011})}\BibitemShut {NoStop}%
\bibitem [{\citenamefont {Bockelmann}\ and\ \citenamefont
  {Bastard}(1990)}]{Bockelmann1990}%
  \BibitemOpen
  \bibfield  {author} {\bibinfo {author} {\bibfnamefont {U.}~\bibnamefont
  {Bockelmann}}\ and\ \bibinfo {author} {\bibfnamefont {G.}~\bibnamefont
  {Bastard}},\ }\href {\doibase 10.1103/PhysRevB.42.8947} {\bibfield  {journal}
  {\bibinfo  {journal} {Physical Review B}\ }\textbf {\bibinfo {volume} {42}},\
  \bibinfo {pages} {8947} (\bibinfo {year} {1990})}\BibitemShut {NoStop}%
\bibitem [{Note100()}]{Note100}%
  \BibitemOpen
  \bibinfo {note} {The coherences between ${\nu }=-1$ and ${\nu }=1$ can be
  neglected if $V {\tau }_{\protect \rm scat}\gg \hbar $, where ${\tau
  }_{\protect \rm scat}$ is the typical scattering time-scale\cite
  {Seetharam2015}.}\BibitemShut {Stop}%
\bibitem [{Note3()}]{Note3}%
  \BibitemOpen
  \bibinfo {note} {We expect our qualitative results to hold also in systems
  with no particle-hole symmetry.}\BibitemShut {Stop}%
\bibitem [{Note4()}]{Note4}%
  \BibitemOpen
  \bibinfo {note} {Slightly better fit could be made by introducing a small
  momentum-dependent shift to the chemical potential. Also, we note that
  although the majority of excitations occupy the band-minima according to, a
  small but finite density of excitations, occupying high quasienergy levels is
  necessary to maintain this distribution.}\BibitemShut {Stop}%
\bibitem [{\citenamefont {Glazman}(1981)}]{Glazman1981}%
  \BibitemOpen
  \bibfield  {author} {\bibinfo {author} {\bibfnamefont {L.~I.}\ \bibnamefont
  {Glazman}},\ }\href@noop {} {\bibfield  {journal} {\bibinfo  {journal} {Zh.
  Eksp. Teor. Fiz.}\ }\textbf {\bibinfo {volume} {80}},\ \bibinfo {pages} {349}
  (\bibinfo {year} {1981})}\BibitemShut {NoStop}%
\bibitem [{\citenamefont {Glazman}(1983)}]{Glazman1983}%
  \BibitemOpen
  \bibfield  {author} {\bibinfo {author} {\bibfnamefont {L.~I.}\ \bibnamefont
  {Glazman}},\ }\href@noop {} {\bibfield  {journal} {\bibinfo  {journal} {Fiz.
  Tekh. Poluprovodn.}\ }\textbf {\bibinfo {volume} {17}},\ \bibinfo {pages}
  {790} (\bibinfo {year} {1983})}\BibitemShut {NoStop}%
\bibitem [{\citenamefont {Galitskii}\ \emph {et~al.}(1969)\citenamefont
  {Galitskii}, \citenamefont {Goreslavskii},\ and\ \citenamefont
  {Elesin}}]{Galitskii1970}%
  \BibitemOpen
  \bibfield  {author} {\bibinfo {author} {\bibfnamefont {V.~M.}\ \bibnamefont
  {Galitskii}}, \bibinfo {author} {\bibfnamefont {S.~P.}\ \bibnamefont
  {Goreslavskii}}, \ and\ \bibinfo {author} {\bibfnamefont {V.~F.}\
  \bibnamefont {Elesin}},\ }\href@noop {} {\bibfield  {journal} {\bibinfo
  {journal} {Zh. Eksp. Teor. Fiz.}\ }\textbf {\bibinfo {volume} {57}},\
  \bibinfo {pages} {207} (\bibinfo {year} {1969})}\BibitemShut {NoStop}%
\bibitem [{Note44()}]{Note44}%
  \BibitemOpen
  \bibinfo {note} {Other choices of ${\Delta }k$ will lead to the same
  power-law dependence.}\BibitemShut {Stop}%
\bibitem [{Note55()}]{Note55}%
  \BibitemOpen
  \bibinfo {note} {Note that for $A=0$, the contribution to the rate coming
  from both the ``vertical'' and ``diagonal'' processes (corresponding to
  momentum transfers $\sim 2k_F$ and $\sim 2k_R$ respectively) yield $\alpha
  =2$ in Eq.~\protect \textup {\hbox {\mathsurround \z@ \protect \normalfont
  (\ignorespaces \ref {eq:AlphaDef}\unskip \@@italiccorr )}}.}\BibitemShut
  {Stop}%
\bibitem [{Note67()}]{Note67}%
  \BibitemOpen
  \bibinfo {note} {The value of ${\alpha }$ depends on the overlap of the
  eigenfunctions. Therefore, it may differ if the system has extra
  symmetries.}\BibitemShut {Stop}%
\bibitem [{\citenamefont {Torres}\ and\ \citenamefont
  {Kunold}(2005)}]{Torres2005a}%
  \BibitemOpen
  \bibfield  {author} {\bibinfo {author} {\bibfnamefont {M.}~\bibnamefont
  {Torres}}\ and\ \bibinfo {author} {\bibfnamefont {A.}~\bibnamefont
  {Kunold}},\ }\href {\doibase 10.1103/PhysRevB.71.115313} {\bibfield
  {journal} {\bibinfo  {journal} {Physical Review B}\ }\textbf {\bibinfo
  {volume} {71}},\ \bibinfo {pages} {115313} (\bibinfo {year}
  {2005})}\BibitemShut {NoStop}%
\bibitem [{Note34()}]{Note34}%
  \BibitemOpen
  \bibinfo {note} {To resolve the transition numerically, in these simulations
  we increased $v_\protect \text {s}$ in the phonon density of states for
  transitions involving large phonon momenta $|q|>k_R$, see Appendix C for
  details.}\BibitemShut {Stop}%
\bibitem [{\citenamefont {Sundaram}\ and\ \citenamefont
  {Mazur}(2002)}]{Sundaram2002}%
  \BibitemOpen
  \bibfield  {author} {\bibinfo {author} {\bibfnamefont {S.~K.}\ \bibnamefont
  {Sundaram}}\ and\ \bibinfo {author} {\bibfnamefont {E.}~\bibnamefont
  {Mazur}},\ }\href {\doibase 10.1038/nmat767} {\bibfield  {journal} {\bibinfo
  {journal} {Nature Materials}\ }\textbf {\bibinfo {volume} {1}},\ \bibinfo
  {pages} {217} (\bibinfo {year} {2002})}\BibitemShut {NoStop}%
\bibitem [{\citenamefont {Ziman}(1956)}]{Ziman1956}%
  \BibitemOpen
  \bibfield  {author} {\bibinfo {author} {\bibfnamefont {J.~M.}\ \bibnamefont
  {Ziman}},\ }\href {\doibase 10.1080/14786435608238092} {\bibfield  {journal}
  {\bibinfo  {journal} {Philosophical Magazine}\ }\textbf {\bibinfo {volume}
  {1}},\ \bibinfo {pages} {191} (\bibinfo {year} {1956})}\BibitemShut {NoStop}%
\end{thebibliography}
\end{document}